\title{\bf \huge Littlest Seesaw }

\author{ 
  {Stephen F.~King\footnote{\tt king@soton.ac.uk}},
\\
  {\small \it Physics and Astronomy, University of Southampton, Southampton, SO17 1BJ, U.K.}\\
       }

\date{}
\documentclass[12pt]{article}
\pdfoutput=1
\usepackage[margin=2cm]{geometry}
\usepackage{color}
\usepackage{graphicx}
\usepackage{amssymb}
\usepackage{amsmath} 
\usepackage{pifont} 
\usepackage{hyperref}
\usepackage[noadjust]{cite}
\usepackage[center,footnotesize,hang]{subfigure}
\usepackage{multirow}

\usepackage{braket}

\usepackage{mathtools}
\usepackage{amssymb}
\usepackage{amsfonts}
\usepackage[footnotesize]{caption}
\usepackage{color}
\usepackage{braket}
\usepackage{cite}
\usepackage{hyperref}
\usepackage{url}
\usepackage{multirow}
\usepackage{relsize}
\usepackage{fullpage}
\usepackage{makecell}
\usepackage{fullpage}

\newcommand{\pmatr}[1]{\begin{pmatrix} #1 \end{pmatrix}}
\newcommand{\simlt}{~\mbox{\smaller\(\lesssim\)}~}
\newcommand{\simgt}{~\mbox{\smaller\(\gtrsim\)}~}

\newcommand{\CP}{$\mathcal{CP}$\,}

\newcommand{\Tr}{\mathrm{Tr}\,}

\def\vev#1{\langle #1 \rangle}

\def\simlt{\stackrel{<}{{}_\sim}}
\def\simgt{\stackrel{>}{{}_\sim}}
\def\be{\begin{equation}}
\def\ee{\end{equation}}
\def\beq{\begin{equation}}
\def\eeq{\end{equation}}
\def\bea{\begin{eqnarray}}
\def\eea{\end{eqnarray}}

\begin{document}

\maketitle

\begin{abstract} 
We propose the Littlest Seesaw (LS) model consisting of just two right-handed neutrinos, where one of them, dominantly responsible for the atmospheric neutrino mass, 
has couplings to $(\nu_e,\nu_{\mu},\nu_{\tau})$ proportional to $(0,1,1)$, while 
the subdominant right-handed neutrino, 
mainly responsible for the solar neutrino mass, has couplings to $(\nu_e,\nu_{\mu},\nu_{\tau})$
proportional to $(1,n,n-2)$. 
This constrained sequential dominance (CSD) model
preserves the first column of the tri-bimaximal (TB) mixing matrix (TM1)
and has a reactor angle
$ \theta_{13} \sim (n-1) \frac{\sqrt{2}}{3}  \frac{m_2}{m_3}$.
This is a generalisation of CSD ($n=1$) which led to TB mixing and arises almost as easily
if $n\geq 1$ is a real number.
We derive exact analytic formulas for the neutrino masses, lepton mixing angles and CP phases
in terms of the four input parameters and discuss exact sum rules. 
We show how CSD ($n=3$)
may arise from vacuum alignment due to 
residual symmetries of $S_4$.
We propose a benchmark model based on $S_4\times Z_3\times Z'_3$, which fixes $n=3$ and the 
leptogenesis phase $\eta = 2\pi/3$, 
leaving only two inputs $m_a$ and $m_b=m_{ee}$ 
describing $\Delta m^2_{31}$,  $\Delta m^2_{21}$ and $U_{PMNS}$. 
The LS model predicts a normal mass hierarchy with a massless
neutrino $m_1=0$ and TM1 atmospheric sum rules. The benchmark LS model additionally predicts:
solar angle $\theta_{12}=34^\circ$,  reactor angle $\theta_{13}=8.7^\circ$,  
atmospheric angle $\theta_{23}=46^\circ$, and 
Dirac phase $\delta_{CP}=-87^{\circ}$.
 \end{abstract}



\section{Introduction}
The discovery of neutrino oscillations, implying mass and mixing, remains one of the greatest discoveries in physics in the last two decades. Although the origin of neutrino mass is presently unknown
(for reviews see e.g.\cite{King:2015aea}), whatever is responsible must be new physics
beyond the Standard Model (BSM).
For example, the leading candidate for neutrino mass and mixing is the seesaw mechanism involving additional
right-handed neutrinos with heavy Majorana masses \cite{seesaw}, providing an elegant 
explanation of the smallness of neutrino mass
\footnote{For a simple introduction to the seesaw mechanism see e.g.\cite{King:2015sfk}.}.

However, in general, the seesaw mechanism typically involves many parameters,
making quantitative predictions of neutrino mass and mixing challenging.
In this respect, the seesaw mechanism offers no more understanding of flavour than the Yukawa couplings
of the SM. Indeed it introduces a new flavour sector associated with right-handed neutrino Majorana masses,
which cannot be directly probed by high energy particle physics experiment.
Clearly a different approach is required to make progress with the new (or nu) Standard Model that
involves the seesaw mechanism. Here we shall make use of the theoretical touchstones of elegance and 
simplicity (which indeed motivate the seesaw mechanism in the first place) to try to allow some 
experimental guidance to inform the high energy seesaw mechanism. If the assumptions we make prove to 
be inconsistent with experiment then we must think again, otherwise the framework of assumptions remains viable.

In this paper, then, we 
focus on natural implementations of the seesaw mechanism, where typically one of the right-handed neutrinos is dominantly responsible
for the atmospheric neutrino mass \cite{King:1998jw}, while a second subdominant right-handed neutrino accounts for the solar neutrino mass \cite{King:1999mb}. This idea of sequential dominance (SD) of right-handed neutrinos
is an elegant hypothesis which, when combined with the assumption
of a zero coupling of the atmospheric neutrino to $\nu_e$, leads to the generic bound $ \theta_{13} \simlt m_2/m_3$ \cite{King:2002nf}, which appears to be approximately saturated according to current measurements of the reactor angle. This bound was derived over a decade before the experimental measurement
of the reactor angle. This success supports the SD approach, and motivates efforts to understand why the
reactor bound is approximately saturated. In order to do this one needs to further constrain the Yukawa couplings
beyond the assumption of a single texture zero as assumed above.

The idea of constrained sequential dominance (CSD) is that the ``atmospheric'' right-handed neutrino 
has couplings to $(\nu_e,\nu_{\mu},\nu_{\tau})$ proportional to $(0,1,1)$,
while the ``solar'' right-handed neutrino has couplings to $(\nu_e,\nu_{\mu},\nu_{\tau})$
proportional to $(1,n,n-2)$ where $n$ is a real number. 
It turns out that such a structure 
preserves the first column of the tri-bimaximal (TB) \cite{Harrison:2002er}
mixing matrix (TM1), leading to the approximate
result $ \theta_{13} \sim (n-1) \frac{\sqrt{2}}{3}  \frac{m_2}{m_3}$,
which we shall derive here from exact results.
This scheme is therefore a generalisation of the original 
CSD($n=1$) \cite{King:2005bj} which led to TB mixing and the prediction
$ \theta_{13} =0 $, which is now excluded. It is also a generalisation of 
CSD($n=2$) \cite{Antusch:2011ic} which predicted $ \theta_{13} \sim \frac{\sqrt{2}}{3}  \frac{m_2}{m_3}$,
which was subsequently proved to be too small when more precise measurements of $\theta_{13}$ were made.
It seems we are third time lucky since CSD($n=3$) \cite{King:2013iva}
predicts $ \theta_{13} \sim 2\frac{\sqrt{2}}{3}  \frac{m_2}{m_3}$ which is in broad agreement 
with current measurements. On the other hand CSD($n=4$) \cite{King:2013xba,King:2013hoa}
predicts $ \theta_{13} \sim \sqrt{2}  \frac{m_2}{m_3}$ appears to be a little too large,
unless a third right-handed neutrino is invoked (unavoidable in Pati-Salam for example \cite{King:2014iia}),
while CSD($n\geq 5$) \cite{Bjorkeroth:2014vha} is excluded.
Recently, CSD($3$) has been exploited 
in Grand Unified Theories (GUTs) based on $SU(5)$ \cite{Bjorkeroth:2015ora,Bjorkeroth:2015tsa} and 
$SO(10)$ \cite{Bjorkeroth:2015uou}. 

The new approach and results in this paper are summarised below:
\begin{itemize}
\item
{The approach here is more general than
previously considered, since we allow $n$ to be a real number, rather than being restricted to the
field of positive integers. The motivation is that the vacuum alignment 
vector $(1,n,n-2)$ is orthogonal to the first column
of the TB matrix $(2,-1,1)$ (which in turn is orthogonal to the second and third TB columns $(1,1,-1)$
and $(0,1,1)$) for any real number $n$, emerges very naturally
as depicted in Fig.\ref{perp}.
This provides a plausible motivation
for considering the vacuum alignment direction $(1,n,n-2)$ for any real number $n$.
We refer to the associated minimal models as the Littlest Seesaw (LS).
The LS with CSD($n$) predicts a normal mass hierarchy with a massless
neutrino $m_1=0$, both testable in the near future. Actually the above predictions also arise in general
two right-handed neutrino models. What distinguishes the LS model from general two right-handed neutrino models are the predictions for the
lepton mixing angles and CP phases as discussed below.}

\item
{For the general case of any real value of $n\geq 1$, for the first time we shall 
derive exact analytic formulas for the neutrino masses, lepton mixing angles and CP phases
(both Dirac and Majorana)
in terms of the four input parameters. 
This is progress since previously only numerical results were used.
We also show that CSD($n$) is subject to the 
TM1 mixing sum rules and no other ones.
From the exact results, which are useful for many purposes but a little lengthy, 
we extract some simple approximations which provide some rough and ready insight into 
what is going on. For example, the approximate result $ \theta_{13} \sim (n-1) \frac{\sqrt{2}}{3}  \frac{m_2}{m_3}$
provides an analytic understanding of why CSD($n\geq 5$) is excluded, which until now has only been a numerical
finding.
}

\item
{We show that the successful case of CSD($3$) arises more naturally from symmetry in the case of $S_4$,
rather than using $A_4$, as was done in previous work
\cite{King:2013iva,King:2013xba,King:2013hoa,King:2014iia,Bjorkeroth:2014vha,Bjorkeroth:2015ora}.
The reason is that both the neutrino scalar vacuum alignments 
$(0,1,1)$ and $(1,3,1)$ preserve residual subgroups of $S_4$ which are not present in $A_4$.
This motivates models based on $S_4$, extending the idea of residual symmetries from 
the confines of two sectors (the charged lepton and neutrino sectors) as is traditionally done in direct models,
to five sectors, two associated with the neutrinos and three with the charged leptons,
as summarised in the starfish shaped diagram in Fig.\ref{starfish}.
}

\item{
Finally we present a benchmark LS model based on 
$S_4\times Z_3\times Z'_3$, with supersymmetric vacuum alignment,
which not only fixes $n=3$ but also the 
leptogenesis phase $\eta = 2\pi/3$, 
leaving only two continuous input masses, yielding two neutrino mass squared splittings and 
the PMNS matrix. A single $Z_3$ factor is required to understand $\eta = 2\pi/3$
as a cube root of unity, while an additional $Z'_3$ is necessary to understand the charged lepton
mass hierarchy and also to help to control the operator structure of the model.
The model provides a simple LS framework for the numerical benchmark predictions:
solar angle $\theta_{12}=34^\circ$,  reactor angle $\theta_{13}=8.7^\circ$,  
atmospheric angle $\theta_{23}=46^\circ$, and 
Dirac phase $\delta_{CP}=-87^{\circ}$,
which are readily testable in forthcoming oscillation experiments.
}
\end{itemize}

The layout of the remainder of the paper is as follows.
In section~\ref{2RHN} we briefly introduce the two right-handed neutrino model
and motivate CSD($n$). 
In section~\ref{TM} we show how CSD($n$) implies TM1 mixing.
In section~\ref{direct} we briefly review the direct and indirect approaches to model building,
based on flavour symmetry.
In section~\ref{indirect} we pursue the indirect approach and show how vacuum alignment for 
CSD($n$) can readily be obtained from the TB vacuum alignments using orthogonality.
In section~\ref{LSmodel} we write down the Lagrangian of the LS model and derive the neutrino mass
matrix from the seesaw mechanism with CSD($n$).
In section~\ref{benchmark} we discuss a numerical benchmark, namely CSD($3$) with
leptogenesis phase $\eta=2\pi/3$ and its connection with the oscillation phase.
In sections~\ref{exactangles}, \ref{exactmasses}, \ref{exactCP} we derive exact analytic formulas
for the angles, masses and CP phases, for the LS model with general CSD($n$) valid for real $n\geq 1$, in terms of the 
four input parameters of the model. In section~\ref{exactsumrules} present the exact 
TM1 atmospheric sum rules, which we argue are the only ones satisfied by the model.
In section~\ref{ra} we focus on the reactor and atmospheric angles and,
starting from the exact results,
derive useful approximate formulae which can provide useful insight.
In section~\ref{S4vac} we show how vacuum alignment for CSD($3$) can arise 
from the residual symmetries of $S_4$, as summarised by the starfish diagram
in Fig.\ref{starfish}.
In sections~\ref{benchmarkmodel} and \ref{benchmarkmodelvacuum}
we present a benchmark LS model based on the discrete group
$S_4\times Z_3\times Z'_3$, with supersymmetric vacuum alignment,
which not only fixes $n=3$ but also the 
leptogenesis phase $\eta = 2\pi/3$, reproducing the parameters of the numerical benchmark.
Section~\ref{conclusion} concludes the paper.
There are two appendices, Appendix~\ref{conventions} on lepton mixing conventions
and Appendix~\ref{S4} on $S_4$.

\section{Seesaw mechanism with two right-handed neutrinos}
\label{2RHN}
The two right-handed neutrino seesaw model was first proposed in \cite{King:1999mb}.
Subsequently
two right-handed neutrino models with two texture zeros were discussed
in \cite{Frampton:2002qc}, however such two texture zero 
models are now phenomenologically excluded \cite{Harigaya:2012bw} for the case of a normal neutrino mass hierarchy considered here. However the two right-handed neutrino model with one texture zero
(actually also suggested in \cite{King:1999mb}), remains viable.

With two right-handed neutrinos, the Dirac mass matrix $m^D$ is, in LR convention,
\begin{equation}
m^D=
\left( \begin{array}{cc}
d & a \\
e & b \\
f & c
\end{array}
\right),\ \ \ \ 
(m^D)^T=
\left( \begin{array}{ccc}
d & e & f\\
a & b& c
\end{array}
\right)
\label{mD0}
\end{equation}

The (diagonal) 
right-handed neutrino heavy Majorana mass matrix $M_{R}$
with rows $(\overline{\nu^{\rm atm}_R}, \overline{\nu^{\rm sol}_R})^T$ and columns $(\nu^{\rm atm}_R, \nu^{\rm sol}_R)$
is,
\begin{equation}
M_{R}=
\left( \begin{array}{cc}
M_{\rm atm} & 0 \\
0 & M_{\rm sol}
\end{array}
\right),\ \ \ \ 
M^{-1}_{R}=
\left( \begin{array}{cc}
M^{-1}_{\rm atm} & 0 \\
0 & M^{-1}_{\rm sol}
\end{array}
\right)
\label{mR0}
\end{equation}

The light effective left-handed Majorana neutrino mass matrix is given by the seesaw formula
\beq
m^{\nu}=-m^DM_{R}^{-1}{m^D}^T,
\label{seesaw}
\eeq
Using the see-saw formula 
dropping the overall minus sign which is physically irrelevant, we find,
by multiplying the matrices in Eqs.\ref{mD0},\ref{mR0},
\begin{equation}
m^{\nu}=m^DM^{-1}_{R}(m^D)^T=
\left( \begin{array}{ccc}
\frac{a^2}{M_{\rm sol}}+ \frac{d^2}{M_{\rm atm}}& \frac{ab}{M_{\rm sol}}+ \frac{de}{M_{\rm atm}} 
& \frac{ac}{M_{\rm sol}}+ \frac{df}{M_{\rm atm}}  \\
\frac{ab}{M_{\rm sol}}+ \frac{de}{M_{\rm atm}} & \frac{b^2}{M_{\rm sol}}+ \frac{e^2}{M_{\rm atm}}  
& \frac{bc}{M_{\rm sol}}+ \frac{ef}{M_{\rm atm}}  \\
\frac{ac}{M_{\rm sol}}+ \frac{df}{M_{\rm atm}} 
& \frac{bc}{M_{\rm sol}}+ \frac{ef}{M_{\rm atm}}
& \frac{c^2}{M_{\rm sol}}+ \frac{f^2}{M_{\rm atm}}  
\end{array}
\right)
\label{2rhn}
\end{equation}

Motivated by the desire to implement the seesaw mechanism in a natural way,
sequential dominance (SD) \cite{King:1998jw,King:1999mb}
assumes that the two right-handed neutrinos
$\nu^{\rm sol}_R$ and $\nu^{\rm atm}_R$ have couplings 
$d\ll e,f$ and 
\begin{equation}
 \frac{(e,f)^2}{M_{\rm atm}} \gg \frac{(a,b,c)^2}{M_{\rm sol}}.
\label{SD0}
\end{equation}
By explicit calculation, using Eq.\ref{2rhn}, one can check that 
in the two right-handed neutrino limit $\det m^{\nu} = 0$.
Since the determinant of a Hermitian matrix is the product of mass eigenvalues 
$$
\det (m^{\nu}m^{{\nu}\dagger}) = m_1^2m_2^2m_3^2,
$$
one may deduce that one of the mass eigenvalues of the complex symmetric matrix above
is zero, which under the SD assumption is the lightest one $m_1=0$
with $m_3\gg m_2$ since the model approximates to a single right-handed neutrino model 
\cite{King:1998jw}.
Hence we see that {\it SD implies a normal neutrino mass hierarchy.}
Including the solar right-handed neutrino as a perturbation, it can be shown that,
for $d=0$, together with the assumption of a dominant atmospheric right-handed neutrino 
in Eq.\ref{SD0}, leads to the approximate results for the solar and atmospheric angles 
\cite{King:1998jw,King:1999mb},
\begin{equation}
\tan \theta_{23}\sim \frac{e}{f}, \ \ \ \ \tan \theta_{12} \sim \frac{\sqrt{2}a}{b-c}.
\label{t120}
\end{equation}
Under the above SD assumption, 
each of the right-handed neutrinos contributes uniquely to a particular physical neutrino mass.
The SD framework above with $d=0$
leads to the relations in Eq.\ref{t120} together with the reactor angle bound \cite{King:2002nf},
\begin{equation}
\theta_{13} \lesssim \frac{m_2}{m_3}
\label{13}
\end{equation}
{\it This result shows that SD allows for large values of the reactor angle, consistent with the 
measured value.} Indeed the measured reactor angle, observed a decade after this 
theoretical bound was derived, approximately saturates the upper limit.
In order to understand why this is so, we must go beyond the SD assumptions
stated so far.

Motivated by the desire to obtain an approximately 
maximal atmospheric angle $\tan \theta_{23}  \sim 1$ and 
trimaximal solar angle 
$\tan \theta_{12} \sim  1/\sqrt{2}$, the results in Eq.\ref{t120} suggest constraining the 
Dirac matrix elements in Eq.\ref{mD0} to take the values $d=0$ with $e=f$ and $b=na$ and also $c = (n-2)a$,
\begin{equation}
	m^D =\pmatr{0 & a \\ e & na \\ e & (n-2)a },	\label{mDn20}
\end{equation}
which, for any positive integer $n$, is referred to as constrained sequential dominance (CSD)
  \cite{King:2005bj,Antusch:2011ic,King:2013iva,King:2013xba,King:2013hoa,King:2014iia,Bjorkeroth:2014vha}.
In section~\ref{ra} we shall show that Eq.\ref{mDn20} also implies,
\beq
\theta_{13} \sim (n-1) \frac{\sqrt{2}}{3}  \frac{m_2}{m_3},
\label{s130}
\eeq
so that the bound in Eq.\ref{13} is approximately saturated for $n\sim 3$.

As already mentioned, we refer to a two right-handed neutrino model in which the Dirac mass matrix in the flavour basis
satisfies Eq.\ref{mDn20}, with $n$ being a real number, as the ``Littlest Seesaw'' or LS model.
The justification 
for this terminology is that it represents the seesaw model with 
the fewest number of parameters consistent with 
current neutrino data. To be precise, in the flavour basis, 
the Dirac mass matrix of the LS model involves two complex parameters $e,a$ plus one real parameter 
$n$. This is fewer than the original 
two right-handed neutrino Dirac mass matrix which involves six complex parameters \cite{King:1999mb}. 
It is also fewer than the two right-handed neutrino model 
in \cite{King:2002nf} which involves five complex parameters due to the single texture zero.
It is even fewer than the minimal right-handed neutrino model in  \cite{Frampton:2002qc}
which involves four complex parameters due to the two texture zeroes. 
It remains to justify the Dirac structure of the LS model in Eq.\ref{mDn20}, and we shall address this
question using symmetry and vacuum alignment in subsequent sections.

\section{Trimaximal Mixing }
\label{TM}

A simple example of lepton mixing which came to dominate the model building community until the measurement of the reactor angle is the tribimaximal (TB) mixing matrix \cite{Harrison:2002er}. 
It predicts zero reactor angle $\theta_{13}=0$,  maximal atmospheric angle $s^2_{23}=1/2$, 
or $\theta_{12}=45^\circ$, and 
a solar mixing angle given by $s_{12}=1/\sqrt{3}$,
i.e. $\theta_{12}\approx 35.3^\circ$. The mixing matrix is given explicitly by
\begin{equation}\label{TB}
U_{\rm TB} =
\left(
\begin{array}{ccc}
\sqrt{\frac{2}{3}} &  \frac{1}{\sqrt{3}}
&  0 \\ - \frac{1}{\sqrt{6}}  & \frac{1}{\sqrt{3}} &  \frac{1}{\sqrt{2}} \\
\frac{1}{\sqrt{6}} & -\frac{1}{\sqrt{3}} &  \frac{1}{\sqrt{2}}    
\end{array}
\right).
\end{equation}%
Unfortunately TB mixing is excluded since it predicts a zero reactor angle.
However CSD in Eq.\ref{mDn20} with two right-handed neutrinos allows a non-zero reactor angle
for $n>1$
and also predicts the lightest physical neutrino mass to be zero, $m_1=0$.
One can also check that the neutrino mass matrix resulting from using Eq.\ref{mDn20} in the 
seesaw formula in Eq.\ref{seesaw},
satisfies
\begin{equation}
m^\nu
\left(
\begin{array}{c}
2 \\
-1\\
1
\end{array}
\right)
=
\left(
\begin{array}{c}
0 \\
0\\
0
\end{array}
\right).
\label{CSD(n)a}
\end{equation}
In other words the column vector $(2,-1,1)^T$
is an eigenvector of $m^\nu $ with a zero eigenvalue, i.e. it is the first column of the PMNS mixing matrix,
corresponding to $m_1=0$,
which means so called TM1 mixing \cite{Xing:2006ms,Albright:2008rp} 
in which the first column of the TB mixing matrix in Eq.\ref{TB}
is preserved, while the other two columns are allowed to differ (in particular the reactor angle will be non-zero
for $n>1$),
\begin{equation}\label{TM1}
U_{\rm TM1} =
\left(
\begin{array}{ccc}
\sqrt{\frac{2}{3}} & -
&  - \\ - \frac{1}{\sqrt{6}}  & - & -  \\
\frac{1}{\sqrt{6}} & - & -  
\end{array}
\right).
\end{equation}%
Interestingly CSD in Eq.\ref{mDn20} with $n=1$  \cite{King:2005bj}
predicts a zero reactor angle and hence TB mixing,
while for $n>1$ it simply predicts the less restrictive TM1 mixing.
Having seen that CSD leads to TB, or more generally TM1 mixing, we now discuss the theoretical origin
of the desired Dirac mass matrix structure in Eq.\ref{mDn20}.

\section{Flavour symmetry: direct versus indirect models}
\label{direct}

Let us expand the neutrino mass matrix in the diagonal charged
lepton basis, assuming exact TB mixing, as
${m^{\nu}_{TB}}=U_{TB}{\rm diag}(m_1, m_2, m_3)U_{TB}^T$ leading
to (absorbing the Majorana phases in $m_i$):
\begin{equation}
\label{mLL} {m^{\nu}_{TB}}= m_1\Phi_1 \Phi_1^T + m_2\Phi_2 \Phi_2^T + m_3\Phi_3 \Phi_3^T
\end{equation}
where 
\beq
\Phi_1^T=\frac{1}{\sqrt{6}}(2,-1,1),\ \ \ \ 
\Phi_2^T=\frac{1}{\sqrt{3}}(1,1,-1), \ \ \ \ 
\Phi_3^T=\frac{1}{\sqrt{2}}(0,1,1),
\label{Phi123}
\eeq
are the respective columns of $U_{TB}$
and $m_i$ are the physical neutrino masses. In the neutrino
flavour basis (i.e. diagonal charged lepton mass basis), it has
been shown that the above TB neutrino mass matrix is invariant
under $S,U$ transformations:
\begin{equation}
{m^{\nu}_{TB}}\,= S {m^{\nu}_{TB}} S^T\,= U {m^{\nu}_{TB}} U^T \ .
\label{S} \end{equation}
A very straightforward argument
\cite{King:2009ap}
shows that this neutrino flavour symmetry group
has only four elements corresponding to Klein's four-group $Z_2^S
\times Z_2^U$. By contrast the diagonal charged lepton mass matrix
(in this basis) satisfies a diagonal phase symmetry $T$. In the case of TB mixing, the
matrices $S,T,U$ form the generators of the group $S_4$ in the
triplet representation, while the $A_4$ subgroup is generated by
$S,T$.

As discussed in \cite{King:2009ap},
the flavour symmetry of the neutrino mass matrix may originate
from two quite distinct classes of models. The class of models,
which we call direct models, are based on a family symmetry such as $S_4$, for example,
where the symmetry of the neutrino mass matrix 
is a remnant of the $S_4$ symmetry of the Lagrangian,
with the generators
$S,U$ preserved in the neutrino sector, while the diagonal generator
$T$ is preserved in the charged lepton sector.
If $U$ is broken but $S$ is preserved, then this leads to TM2 mixing with the second column of the TB
mixing matrix being preserved. However if the combination $SU$ is preserved then this corresponds to 
TM1 mixing with the first column of the TB
mixing matrix being preserved \cite{Luhn:2013vna}.
Of course, the $S_4$ symmetry is completely broken in the full lepton Lagrangian including both
neutrino and charged lepton sectors.

In an alternative class of models, which we call indirect
models, the family symmetry is already completely
broken in the neutrino sector, where the observed neutrino flavour
symmetry $Z_2^S \times Z_2^U$ emerges as an
accidental symmetry.
However the structure of the Dirac mass matrix is controlled by vacuum alignment 
in the flavour symmetry breaking sector, as discussed in the next section.
The indirect models are arguably more natural than the direct models,
especially for $m_1=0$, since each column
of the Dirac mass matrix corresponds to a different symmetry breaking VEV and each contribution to the 
seesaw mechanism corresponds to a different right-handed neutrino mass, enabling 
mass hierarchies to naturally emerge. Thus a strong mass hierarchy $m_1\ll m_2 < m_3$ 
would seem to favour indirect models over direct models, so we pursue this possibility in the following.

\section{Indirect approach and vacuum alignment}
\label{indirect}

\begin{figure}[htb]
\centering
\includegraphics[width=0.5\textwidth]{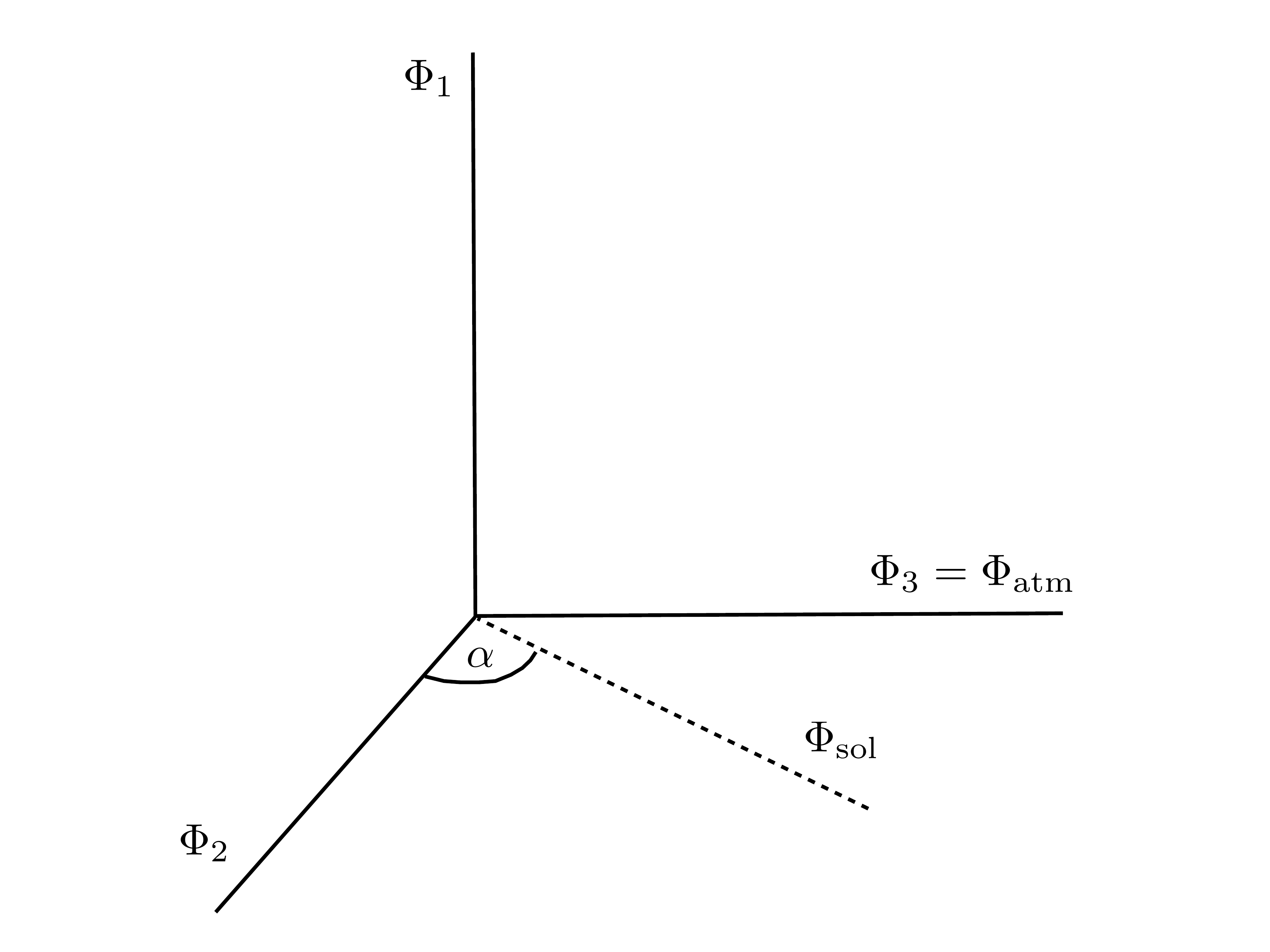}
\caption{\label{perp}\small{The mutually orthogonal vacuum alignments $\Phi_i$ 
in Eq.\ref{Phi123}
used for TB mixing.
The alignment vector $\Phi_{\rm sol}$ is orthogonal to $ \Phi_1$ and hence is in the plane defined by
 $ \Phi_2$ and  $ \Phi_3= \Phi_{\rm atm}$.
 Note that the vectors $\Phi_{\rm sol}$ and $\Phi_{\rm atm}$ in Eq.\ref{Phias} are not orthogonal 
 for a general value of $n$, so any seesaw model based on these alignments 
 will violate form dominance. 
}}
\end{figure}

The basic idea of the indirect approach is to effectively promote the 
columns of the Dirac mass matrix to fields which transform as triplets under the flavour symmetry. We assume that the Dirac
mass matrix can be written as $m_D= (a\Phi_{\rm atm},b\Phi_{\rm sol},c\Phi_{\rm dec})$ 
where the columns are proportional to triplet Higgs scalar fields with 
particular vacuum alignments and $a,b,c$ are three constants of proportionality. Working 
in the diagonal right-handed neutrino mass basis, the seesaw formula
gives,
\begin{equation}
\label{mLLCSD} {m^{\nu}}=
  a^2\frac{\Phi_{\rm atm}\Phi_{\rm atm}^T}{M_{\rm atm}}
+ b^2\frac{\Phi_{\rm sol}\Phi_{\rm sol}^T}{M_{\rm sol}}
+ c^2\frac{\Phi_{\rm dec}\Phi_{\rm dec}^T}{M_{\rm dec}},
\end{equation}
By comparing Eq.\ref{mLLCSD} to the TB form in Eq.\ref{mLL} it is
clear that TB mixing will be achieved if $\Phi_{\rm atm}\propto \Phi_3$ and 
$\Phi_{\rm sol}\propto \Phi_2$ and $\Phi_{\rm dec}\propto \Phi_1$, with each of $m_{3,2,1}$
originating from a particular right-handed neutrino.
The case where the columns of the Dirac mass matrix are proportional to the columns of the
PMNS matrix, the columns being therefore mutually orthogonal,
is referred to as form dominance (FD) \cite{Chen:2009um}. 
The resulting ${m^{\nu}}$ is form diagonalizable.
Each column of the Dirac mass matrix 
arises from a separate flavon VEV, so the mechanism is very natural, especially for the case of a strong mass
hierarchy.
Note that for $m_1\ll m_2 < m_3$ the precise form of $\Phi_{\rm dec}$ becomes
irrelevant and for $m_1=0$ we can simply drop the last term and the model reduces to a two right-handed 
neutrino model.

Within this framework, the general CSD 
Dirac mass matrix structure in Eq.\ref{mDn20} corresponds to there being some 
Higgs triplets which can be aligned in the directions,
\beq
\Phi_{\rm atm}^T\propto (0,1,1), \ \ \ \ 
\Phi_{\rm sol}^T\propto (1,n,n-2), 
\label{Phias}
\eeq
The first vacuum alignment $\Phi_{\rm atm}$ in Eq.\ref{Phias} is just the TB direction 
$\Phi_3^T$ in Eq.\ref{Phi123}.
The second vacuum alignment $\Phi_{\rm sol}$ in Eq.\ref{Phias} 
can be easily obtained since 
the direction $(1,n,n-2)$ is orthogonal to the TB vacuum alignment $\Phi_1^T$ in Eq.\ref{Phi123}.

For example, in a supersymmetric theory,
the aligning superpotential should contain the following terms, enforced by suitable discrete $Z_n$ symmetries,
\beq
O_{ij}\phi_i \phi_j + O_{\rm sol}\phi_{\rm sol} \phi_1
\label{orthog}
\eeq
where the terms proportional to the singlets $O_{ij}$ and $O_{\rm sol}$ ensure that the real $S_4$ triplets 
are aligned in mutually orthogonal directions, $\Phi_i \perp \Phi_j $ and $\Phi_{\rm sol} \perp \Phi_1$
as depicted in Fig.~\ref{perp}.
From Eqs.\ref{Phi123},\ref{Phias} we can write 
 $\Phi_{\rm sol}= \Phi_2 \cos \alpha  + \Phi_3 \sin \alpha $ where $\tan \alpha = 2(n-1)/3$
and $\alpha$ is the angle between $\Phi_{\rm sol}$ and $\Phi_2$, as shown in Fig.~\ref{perp}. 
$\Phi_{\rm sol}$ is parallel to $\Phi_2$ for $n=1$, while it increasingly tends towards 
 the $ \Phi_3= \Phi_{\rm atm}$ alignment as $n$ is increased.

\section{The Littlest Seesaw}
\label{LSmodel}
The littlest seesaw  (LS) model consists of 
the three families of electroweak lepton doublets $L$ unified into a single triplet of the flavour symmetry,
while the two right-handed neutrinos $\nu^{\rm atm}_R$ and $\nu^{\rm sol}_R$ are singlets.
The LS Lagrangian in the neutrino sector takes the form,
\beq
{\cal L}=-y_{\rm atm}\bar{L}.\phi_{\rm atm}\nu_R^{\rm atm}-y_{\rm sol}\bar{L}.\phi_{\rm sol}\nu_R^{\rm sol} 
- \frac{1}{2}M_{\rm atm}\bar{\nu^c_R}^{\rm atm}\nu_R^{\rm atm}
- \frac{1}{2}M_{\rm sol}\bar{\nu^c_R}^{\rm sol}\nu_R^{\rm sol}
+ {H.c.}\; .
\label{LS}
\eeq
which may be enforced by suitable discrete $Z_3$ symmetries, 
as discussed in section~\ref{benchmarkmodel}.
Here $\phi_{\rm sol}$ and $\phi_{\rm atm}$ may be interpreted as either Higgs fields, 
which transform as triplets of the flavour symmetry
with the alignments in Eq.\ref{Phias}, or as combinations of a single Higgs electroweak doublet together with 
triplet flavons with these vacuum alignments. Note that, in Eq.\ref{LS}, $\phi_{\rm sol}$ and $\phi_{\rm atm}$
represent fields, whereas in Eq.\ref{Phias} $\Phi_{\rm sol}$ and $\Phi_{\rm atm}$
refer to the VEVs of those fields.

In the diagonal charged lepton and right-handed neutrino mass basis,
when the fields $\Phi_{\rm sol}$ and $\Phi_{\rm atm}$ in Eq.\ref{LS} are replaced by their VEVs
in Eq.\ref{Phias}, this reproduces the Dirac mass matrix in Eq.\ref{mDn20} \cite{King:2013iva}
and its transpose:
\begin{equation}
	m^D = \pmatr{0 & a \\ e & na \\ e & (n-2)a },	\ \ \ \ 
	{m^D}^T = \pmatr{0 & e & e \\ a & na & (n-2)a },
	\label{mDn}
\end{equation}
which defines the LS model, where we regard $n$ as a real continuous parameter, later arguing that it may take 
simple integer values.
The (diagonal) 
right-handed neutrino heavy Majorana mass matrix $M_{R}$
with rows $(\bar{\nu^c_R}^{\rm atm}, \bar{\nu^c_R}^{\rm sol})^T$ and columns $(\nu^{\rm atm}_R, \nu^{\rm sol}_R)$
is,
\begin{equation}
M_{R}=
\left( \begin{array}{cc}
M_{\rm atm} & 0 \\
0 & M_{\rm sol}
\end{array}
\right),\ \ \ \ 
M^{-1}_{R}=
\left( \begin{array}{cc}
M^{-1}_{\rm atm} & 0 \\
0 & M^{-1}_{\rm sol}
\end{array}
\right)
\label{mR}
\end{equation}

The low energy effective Majorana neutrino mass matrix is given by the seesaw formula
\beq
m^{\nu}=-m^DM_{R}^{-1}{m^D}^T,
\label{seesawp}
\eeq
which, after multiplying the matrices in Eqs.\ref{mDn}, \ref{mR}, for 
a suitable choice of physically irrelevant overall phase, gives
\begin{equation}
	m^\nu = m_a 
	\left(
\begin{array}{ccc}
	0&0&0\\0&1&1\\0&1&1 
	\end{array}
\right)
	+ m_b e^{i\eta} 
	\left(
\begin{array}{ccc}
	1&n&(n-2)\\n&n^2&n(n-2)\\(n-2)&n(n-2)&(n-2)^2
	\end{array}
\right),
	\label{eq:mnu2}
\end{equation}
where $\eta$ is the only physically important phase, which depends on the relative phase between the first and second column of the Dirac mass matrix, $\arg (a/e)$ and 
$m_a=\frac{|e|^2}{M_{\rm atm}}$ and $m_b=\frac{|a|^2}{M_{\rm sol}}$.
This can be thought of as the minimal (two right-handed neutrino) predictive seesaw model since
only four real parameters $m_a, m_b, n, \eta$ describe the entire neutrino sector (three neutrino masses
as well as the PMNS matrix, in the diagonal charged lepton mass basis).
As we shall see in the next section, $\eta$ is identified with the leptogenesis phase, while 
$m_b$ is identified with the neutrinoless double beta decay parameter $m_{ee}$.

\section{A numerical benchmark: CSD(3) with $\eta = 2\pi/3$}
\label{benchmark}
We now illustrate the success of the scheme by presenting numerical results for 
the neutrino mass matrix in Eq.\ref{eq:mnu2} for the particular choice 
of input parameters, namely $n=3$ and $\eta = 2\pi/3$,
\footnote{Note that the seesaw mechanism results in a light effective Majorana mass matrix 
given by the Lagrangian 
${\cal L_{\rm eff}}=- \frac{1}{2}\overline{\nu_L} m^{\nu} \nu_{L}^c + H.c.$.
This corresponds to the convention of Appendix~\ref{conventions}.}
\begin{equation}
	m^\nu = m_a 
	\left(
\begin{array}{ccc}
	0&0&0\\0&1&1\\0&1&1 
	\end{array}
\right)
	+ m_b e^{i 2\pi/3} 
	\left(
\begin{array}{ccc}
	1&3&1\\3&9&3\\1&3&1
	\end{array}
\right).
	\label{eq:mnu2p}
\end{equation}
This numerical benchmark was first presented in \cite{Bjorkeroth:2015ora,Bjorkeroth:2015tsa}.
In section~\ref{benchmarkmodel} we will propose a simple LS
model which provides a theoretical justification for this choice of parameters.

In Table \ref{tab:model} we compare the above numerical benchmark
resulting from the neutrino mass matrix in Eq.\ref{eq:mnu2p} to the global best fit values 
from \cite{Gonzalez-Garcia:2014bfa} (setting $m_1=0$). 
The agreement between CSD(3) and data is 
within about one sigma for all the parameters,
with similar agreement for the other global fits \cite{Capozzi:2013csa,Forero:2014bxa}.

\begin{table}[ht]
\renewcommand{\arraystretch}{1.2}
\centering
\begin{tabular}{| c c c | c c c c c c c |}
\hline
\rule{0pt}{4ex}%
\makecell{$m_a$ \\ {\scriptsize (meV)}} & \makecell{$m_b$ \\ {\scriptsize (meV)}} & 
\makecell{$\eta$  \\ {\scriptsize (rad)}}  	& \makecell{$\theta_{12}$ \\ {\scriptsize ($^{\circ}$)}} & \makecell{$\theta_{13}$ \\ {\scriptsize ($^{\circ}$)}}  & \makecell{$\theta_{23}$ \\ {\scriptsize ($^{\circ}$)}} & \makecell{$\delta_{\mathrm{CP}}$ \\ {\scriptsize ($^{\circ}$)}} & \makecell{$m_1$ \\ {\scriptsize (meV)}}
& \makecell{$m_2$ \\ {\scriptsize (meV)}} & \makecell{$m_3$ \\ {\scriptsize (meV)}} \\ [2ex] \hline 
\rule{0pt}{4ex}%
26.57		& 2.684		& $ \dfrac{2\pi}{3} $	& 34.3		& 8.67		& 45.8		& -86.7		& 0 & 8.59		& 49.8 \\[1.7ex]
\hline
Value & from & \cite{Gonzalez-Garcia:2014bfa} & 33.48$^{+0.78}_{-0.75}$ & 8.50$^{+0.20}_{-0.21}$ &
42.3$^{+3.0}_{-1.6}$  &  -54$^{+39}_{-70}$ & 0 & 8.66$\pm 0.10$ & 49.57$\pm 0.47$ \\
\hline
\end{tabular}
\caption{Parameters and predictions for CSD(3) with a fixed phase $ \eta = 2\pi/3 $
from \cite{Bjorkeroth:2015ora}.
In addition we predict $\beta = 71.9^{\circ}$ which is not shown in the Table  
since the neutrinoless double beta decay parameter
is $m_{ee}=m_b= 2.684$ meV for the above parameter set which is practically impossible
to measure in the forseeable future.
These predictions may be compared to the global best fit values 
from \cite{Gonzalez-Garcia:2014bfa} (for $m_1=0$), given on the last line.}
\label{tab:model}
\end{table}

Using the results in Table \ref{tab:model}, the baryon asymmetry of the Universe
(BAU) resulting from $N_1 = N_{\rm atm}$ leptogenesis was estimated for this model \cite{Bjorkeroth:2015tsa}:
\begin{equation}
	Y_B \approx 2.5 \times 10^{-11}\sin \eta \left[\frac{M_1}{10^{10} ~\mathrm{GeV}} \right].
\label{BAU}
\end{equation}
Using $\eta = 2\pi/3$ and the observed value of $ Y_B $ fixes the lightest right-handed neutrino mass:
\begin{equation}
	M_1 = M_{\rm atm}  \approx 3.9 \times 10^{10} ~\mathrm{GeV}.
\end{equation}
The phase $\eta$ determines the BAU via leptogenesis in Eq.\ref{BAU}.
In fact it controls the entire PMNS matrix, including all the lepton mixing angles as well as all low energy \CP violation.
The leptogenesis phase $\eta$ is therefore the source of all \CP violation arising from this model,
including \CP violation in neutrino oscillations and in leptogenesis.
There is a direct link between measurable and cosmological \CP violation in this model and 
a correlation between the sign of the BAU and the sign of 
low energy leptonic \CP violation. The
leptogenesis phase is fixed to be $\eta = 2\pi/3$ which leads to the observed excess of matter over antimatter
for $M_1 \approx 4.10^{10}$ GeV, yielding 
an observable neutrino oscillation phase $\delta_{\mathrm{CP}} \approx -\pi/2$.

\section{Exact analytic results for lepton mixing angles}
\label{exactangles}
We would like to understand the numerical success of the neutrino mass matrix analytically.
In the following sections we shall derive exact analytic 
results for neutrino masses and PMNS parameters, for real continuous $n$,
corresponding to the physical (light effective left-handed Majorana) neutrino mass matrix,
in the diagonal charged lepton mass
basis in Eq.\ref{eq:mnu2}, which we reproduce below,
\begin{equation}
	m^\nu = m_a 
	\left(
\begin{array}{ccc}
	0&0&0\\0&1&1\\0&1&1 
	\end{array}
\right)
	+ m_b e^{i\eta} 
	\left(
\begin{array}{ccc}
	1&n&n-2\\n&n^2&n(n-2)\\1&n(n-2)&(n-2)^2
	\end{array}
\right).
	\label{eq:mnu2pp}
\end{equation}

Since this yields TM1 mixing as discussed above, it can be block diagonalised by the TB mixing matrix,
\begin{equation}
	m^\nu_{block} = U_{\rm TB}^T m^\nu U_{\rm TB} =
\left(
\begin{array}{ccc}
	0&0&0\\0&x & y\\0& y &
	z
	\end{array}
\right)	
\label{eq:mnu3}
\end{equation}
where we find,
\beq
x= 3m_b e^{i\eta} , \ \ \ \ 
y=  \sqrt{6}m_b e^{i\eta} (n-1),\ \ \ \ 
z= |z|e^{i\phi_z}= 2[ m_a+ m_b e^{i\eta} (n-1)^2 ]
\label{xyz}
\eeq
It only remains to put $m^\nu_{\rm block} $ into diagonal form, with real positive masses,
which can be done exactly analytically of course,
since this is just effectively a two by two complex symmetric matrix,
\beq
 U_{\rm block}^T m^\nu_{\rm block} U_{\rm block} =P_{3\nu}^*R_{23\nu}^TP_{2\nu}^*m^\nu_{\rm block} P_{2\nu}^*R_{23\nu}P_{3\nu}^*=m^{\nu}_{\rm diag}={\rm diag}(0,m_2,m_3),
\label{diagnu}
\eeq
where 
\beq
P_{2\nu}=\left(
\begin{array}{ccc}
	1&0&0\\
	0&e^{i\phi^{\nu}_2}  & 0\\
	0& 0 & e^{i\phi^{\nu}_3} 
	\end{array}
\right)
\label{P2}
\eeq
\beq
P_{3\nu}=\left(
\begin{array}{ccc}
	e^{i\omega^{\nu}_1}&0&0\\
	0&e^{i\omega^{\nu}_2}  & 0\\
	0& 0 & e^{i\omega^{\nu}_3} 
	\end{array}
\right)
\label{P3}
\eeq
and 
\beq
R_{23\nu}=\left(
\begin{array}{ccc}
	1&0&0\\
	0&\cos \theta^{\nu}_{23}  & \sin \theta^{\nu}_{23}\\
	0& -\sin \theta^{\nu}_{23} & \cos \theta^{\nu}_{23}
	\end{array}
\right)
\equiv
	\left(
\begin{array}{ccc}
	1&0&0\\
	0&c^{\nu}_{23}  & s^{\nu}_{23}\\
	0& -s^{\nu}_{23} & c^{\nu}_{23}
	\end{array}
\right)
\label{R23}
\eeq
where the angle $\theta^{\nu}_{23}$ is given exactly by,
\beq
t\equiv \tan 2 \theta^{\nu}_{23} =   \frac{2 |y|}{|z|\cos (A - B) -|x| \cos B}  
\label{t}
\eeq
where 
\beq
\tan B =  \tan (\phi^{\nu}_3-\phi^{\nu}_2  ) = \frac{ |z|\sin A}{|x|+|z| \cos A}  
\label{B}
\eeq
where $x,y,z$ were defined in terms of input parameters in Eq.\ref{xyz} and
\beq
A= \phi_z - \eta  = \arg [m_a+ m_b e^{i\eta} (n-1)^2 ] - \eta .\label{A}
\eeq
Recall from Eq.\ref{eq:DiagMe},
 \begin{eqnarray}\label{eq:DiagMnu}
V_{\nu_L} \,m^{\nu}\,V^T_{\nu_L} = m^{\nu}_{\rm diag}=
\mbox{diag}(m_1,m_2,m_3) =
\mbox{diag}(0,m_2,m_3).
\label{mLLnu}
\end{eqnarray}
From Eqs.\ref{eq:mnu3},\ref{diagnu},\ref{eq:DiagMnu} we identify,
\beq
V_{\nu_L} = U_{\rm block}^T U_{\rm TB}^T = P_{3\nu}^*R_{23\nu}^TP_{2\nu}^* U_{\rm TB}^T, 
\ \ 
V_{\nu_L}^\dagger =U_{\rm TB} U_{\rm block}^* = U_{\rm TB}  P_{2\nu}R_{23\nu}P_{3\nu}.
\label{Vnu}
\eeq
Explicitly we find from Eq.\ref{Vnu}, using Eqs.\ref{TB},\ref{P2},\ref{P3},\ref{R23},
\bea
 \label{eq:Vnumatrix}
V_{\nu_L}^\dagger =
\left(
\begin{array}{ccc}
\sqrt{\frac{2}{3}} &  \frac{e^{i\phi^{\nu}_2}}{\sqrt{3}}c^{\nu}_{23} &   \frac{e^{i\phi^{\nu}_2}}{\sqrt{3}}s^{\nu}_{23} \\ 
- \frac{1}{\sqrt{6}} 
& \frac{e^{i\phi^{\nu}_2}}{\sqrt{3}}c^{\nu}_{23}- \frac{e^{i\phi^{\nu}_3}}{\sqrt{2}}s^{\nu}_{23}
&  \frac{e^{i\phi^{\nu}_3}}{\sqrt{2}}c^{\nu}_{23}+ \frac{e^{i\phi^{\nu}_2}}{\sqrt{3}}s^{\nu}_{23} \\
\frac{1}{\sqrt{6}} & 
 -\frac{e^{i\phi^{\nu}_2}}{\sqrt{3}}c^{\nu}_{23}- \frac{e^{i\phi^{\nu}_3}}{\sqrt{2}}s^{\nu}_{23}
 & \frac{e^{i\phi^{\nu}_3}}{\sqrt{2}}c^{\nu}_{23}- \frac{e^{i\phi^{\nu}_2}}{\sqrt{3}}s^{\nu}_{23}  
\end{array}
\right)
\left(
\begin{array}{ccc}
	e^{i\omega^{\nu}_1}&0&0\\
	0&e^{i\omega^{\nu}_2}  & 0\\
	0& 0 & e^{i\omega^{\nu}_3} 
	\end{array}
\right).
    \eea
    which takes the trimaximal form of Eq.\ref{TM1}.
    Recall from Eqs.\ref{Eq:PMNS_Definition},\ref{PE},\ref{Vnu}, the PMNS matrix is given by,
\begin{eqnarray}\label{Eq:PMNS_Definition2}
U
= V_{E_{L}} V^\dagger_{\nu_{L}}
\end{eqnarray}
Writing $(V^\dagger_{\nu_{L}})_{ij}=e^{i\rho^{\nu}_{ij} }|(V^\dagger_{\nu_{L}})_{ij}|$,
and introducing two charged lepton phases, $\phi_{\mu}$ and $\phi_{\tau}$,
\bea
 \label{eq:Umatrix}
U =
\left(
\begin{array}{ccc}
\sqrt{\frac{2}{3}} &  \frac{1}{\sqrt{3}}c^{\nu}_{23} &   
\frac{1}{\sqrt{3}}s^{\nu}_{23} \\ 
e^{i\phi_{\mu}}\frac{-1}{\sqrt{6}}  
&e^{i\phi_{\mu}}( \frac{1}{\sqrt{3}}c^{\nu}_{23}- \frac{e^{iB}}{\sqrt{2}}s^{\nu}_{23})
& e^{i(\phi_{\mu}+\rho^{\nu}_{23} )}| \frac{e^{iB}}{\sqrt{2}}c^{\nu}_{23}+ \frac{1}{\sqrt{3}}s^{\nu}_{23} | 
\\
e^{i\phi_{\tau}}\frac{1}{\sqrt{6}}  & 
e^{i\phi_{\tau}}( -\frac{1}{\sqrt{3}}c^{\nu}_{23}- \frac{e^{iB}}{\sqrt{2}}s^{\nu}_{23})
 & e^{i(\phi_{\tau}+\rho^{\nu}_{33} )} | \frac{e^{iB}}{\sqrt{2}}c^{\nu}_{23}- \frac{1}{\sqrt{3}}s^{\nu}_{23} | 
\end{array}
\right)
\left(\begin{array}{ccc}
e^{i\omega^{\nu}_1} & 0 & 0 \\
0 & e^{i(\phi^{\nu}_2+\omega^{\nu}_2)} & 0\\
0 & 0 & e^{i(\phi^{\nu}_2+\omega^{\nu}_3)} \\
\end{array}\right)
    \eea    
        which can be compared to the PDG parameterisation in Eq.\ref{DM},\ref{V},\ref{Maj},
    \bea
 \label{eq:UmatrixPDG}
U =
\left(\begin{array}{ccc}
    c_{12} c_{13}
    & s_{12} c_{13}
    & s_{13} e^{-i\delta}
    \\
    - s_{12} c_{23} - c_{12} s_{13} s_{23} e^{i\delta}
    & \hphantom{+} c_{12} c_{23} - s_{12} s_{13} s_{23}
    e^{i\delta}
    & c_{13} s_{23} \hspace*{5.5mm}
    \\
    \hphantom{+} s_{12} s_{23} - c_{12} s_{13} c_{23} e^{i\delta}
    & - c_{12} s_{23} - s_{12} s_{13} c_{23} e^{i\delta}
    & c_{13} c_{23} \hspace*{5.5mm}
    \end{array}\right)
    \left(\begin{array}{ccc}
e^{i\frac{\beta_1}{2}} & 0 & 0 \\
0 & e^{i\frac{\beta_2}{2}} & 0\\
0 & 0 & 1 \\
\end{array}\right)
    \eea
from which comparison we identify the physical PMNS lepton mixing angles by the exact expressions
\bea
\sin \theta_{13} &=& \frac{1}{\sqrt{3}}s^{\nu}_{23} =  \frac{1}{\sqrt{6}}\left(1-\sqrt{\frac{1 }{1+t^2}}       \right)^{1/2}
\label{s13}\\
\tan \theta_{12} &=& \frac{1}{\sqrt{2}}c^{\nu}_{23} = \frac{1}{\sqrt{2}}\left( 1- 3 \sin^2 \theta_{13}    \right)^{1/2}
\label{t12}\\
\tan \theta_{23} &=& \frac{| \frac{e^{iB}}{\sqrt{2}}c^{\nu}_{23}+ \frac{1}{\sqrt{3}}s^{\nu}_{23} |}
{| \frac{e^{iB}}{\sqrt{2}}c^{\nu}_{23}- \frac{1}{\sqrt{3}}s^{\nu}_{23} | } 
= \frac{|1+\epsilon^{\nu}_{23}|}{|1-\epsilon^{\nu}_{23}|}
\label{t23}
\eea
where we have selected the negative sign for the square root in parentheses, applicable for the physical range of parameters, and defined 
\bea
\epsilon^{\nu}_{23} & \equiv & \sqrt{\frac{2}{3}}\tan \theta^{\nu}_{23}e^{-iB} = \sqrt{\frac{2}{3}}t^{-1} 
\left[\sqrt{1+t^2 } -1 \right] e^{-iB}
\label{epsnu}
\eea
expressing the results in terms of $t\equiv \tan 2 \theta^{\nu}_{23}$ and $B= \phi^{\nu}_3-\phi^{\nu}_2$ which were given in terms of input parameters in Eqs.\ref{t},\ref{B},\ref{A}.
The solar angle $\tan \theta_{12}$ approximately takes the TB value of $1/\sqrt{2}$, to first order 
in $\sin \theta_{13}$. The atmospheric angle $\tan \theta_{23}$ is maximal when 
$B=\pm \pi/2$ since then $|1+\epsilon^{\nu}_{23}|$ is equal to $|1-\epsilon^{\nu}_{23}|$.

\section{Exact analytic results for neutrino masses}
\label{exactmasses}

The neutrino masses can be calculated from the block diagonal form of the neutrino mass matrix in 
Eq.\ref{eq:mnu3} which is diagonalised with real postive mass eigenvalues $m_2,m_3$ as in 
Eq.\ref{diagnu}. After forming the Hermitian combination (in terms of $x,y,z$ in Eq.\ref{xyz}),
\beq
H^\nu_{\rm block} = m^\nu_{\rm block} m^{\nu \dag}_{\rm block} = \left(
\begin{array}{ccc}
	0&0&0\\0&|x|^2+ |y|^2& |x| |y| +  |y|  e^{i\eta} z^*\\0&  |x| |y| +  |y|  e^{-i\eta} z &
	|y|^2+ |z|^2
	\end{array}
\right)
\label{H}	
\eeq
we diagonalise it by,
\beq
   U_{\rm block}^T H^\nu_{\rm block} U_{\rm block}^* = 
{\rm diag}(0,m^2_2,m^2_3).
\label{Hdiag}
\eeq
Then, by taking the Trace (T) and Determinant (D) of Eq.\ref{Hdiag}, using Eq.\ref{H}, we find
\bea
m_2^2+m_3^2 &=& T \equiv  |x|^2+ 2|y|^2 +|z|^2\\
m_2^2m_3^2 &=& D \equiv |x|^2|z|^2 +|y|^4   - 2 |x||y|^2|z|\cos A
\eea
from which we extract the exact results for the neutrino masses,
\bea
m_3^2 & = & \frac{1}{2}T +   \frac{1}{2} \sqrt{T^2-4D}\label{m3}\\
m_2^2 & =  & D / m_3^2 \label{m2}\\
m_1^2 & =  & 0 \label{m1}
\eea
where we have selected the positive sign for the square root which is applicable for $m_3^2 > m_2^2$.
Furthermore, since $m_3^2 \gg m_2^2$ (recall $m_3^2/m_2^2\approx 30$ ) we may approximate,
\bea
m_3^2 &\approx &T=  |x|^2+ 2|y|^2 +|z|^2 \label{m3approx}\\
m_2^2 &\approx & D/T = \frac{|x|^2|z|^2 +|y|^4   - 2 |x||y|^2|z|\cos A}{|x|^2+ 2|y|^2 +|z|^2} \label{m2approx}
\eea
The sequential dominance (SD) approximation that the atmospheric right-handed neutrino
dominates over the solar right-handed neutrino contribution to the seesaw mass matrix 
implies that $m_a\gg m_b$ and $|z|\gg |x|,|y|$ leading to 
\bea
m_3^2 &\approx & |z|^2 \ \ \ \ \longrightarrow \ \ \ \ m_3\approx 2m_a \label{m3SD}\\
m_2^2 &\approx & |x|^2 \ \ \ \ \longrightarrow \ \ \ \ m_2\approx 3m_b \label{m2SD}
\eea
using the definitions in Eq.\ref{xyz}. 
The results in Eqs.\ref{m3SD} and \ref{m2SD} are certainly very simple, but how accurate are they?
For the CSD(3) numerical benchmark $m_a\approx 27$ meV and $m_b\approx 2.7$ meV gives
$m_3\approx 50$ meV, $m_2\approx 8.6$ meV. We find
$|z|\approx 46$ meV , $|x|\approx 8.0$ meV (and $|y|\approx 13$ meV) which is a reasonable approximation.
The SD approximations in Eqs.\ref{m3SD} and \ref{m2SD} give,
$m_3\approx 2m_a \approx 54$ meV and $m_2\approx 3m_b \approx 8.1$ meV,
accurate to say 10\%. The approximate results in Eqs.\ref{m3approx},\ref{m2approx} are more accurate to say 3\%, 
with the results in Eqs.\ref{m3},\ref{m2} being of course exact.

The SD approximation in Eqs.\ref{m3SD}, \ref{m2SD} is both insightful and useful, since
two of the three input parameters, namely $m_a$ and $m_b$, are immediately fixed by the two physical
neutrino masses $m_3$ and $m_2$, which, for $m_1=0$, are identified as the square roots of the measured
mass squared differences $\Delta m_{31}^2$ and $\Delta m_{21}^2$. This leaves, in the SD approximation,
the only remaining parameters to be $n$ and the phase $\eta$, which, together, determine the entire PMNS
mixing matrix (3 angles, and 2 phases). For example if $n=3$ and $\eta = 2\pi /3$ were determined by some model, then the PMNS matrix would be determined uniquely, without any freedom, in the SD approximation.
When searching for a best fit solution, the SD approximation in in Eqs.\ref{m3SD}, \ref{m2SD} 
is useful as a first approximation which enables the 
parameters $m_a$ and $m_b$ to be approximately determined by $\Delta m_{31}^2$ and $\Delta m_{21}^2$
since this may then be used 
as a starting point around which a numerical minimisation package can be run using the exact results for the neutrino masses in Eqs.\ref{m3},\ref{m2} together with the exact results for the lepton mixing angles in Eqs.\ref{s13},\ref{t12},\ref{t23}.

\section{Exact analytic results for CP violation}
\label{exactCP}

In this model the Majorana phase $\beta_1$ is unphysical since $m_1=0$ so the only physical Majorana phase is $\beta \equiv \beta_2$.
By comparing Eqs.\ref{eq:Umatrix}, \ref{eq:UmatrixPDG} the physical PMNS phases are then identified as 
\bea
\delta &=&  \rho^{\nu}_{23} +\phi_{\mu}  =  \rho^{\nu}_{33} +\phi_{\tau} \label{delta} \\
\frac{\beta}{2} & = & \omega^{\nu}_2 - \omega^{\nu}_3 -\delta
\label{beta}
\eea
Extracting the value of the physical phases $\delta$ and  $\beta$ in terms of input parameters
is rather cumbersome and it is better to use the Jarlskog and Majorana invariants in order to do this.

\subsection{The Jarlskog Invariant}

The Jarlskog invariant $J$ \cite{Jarlskog:1985ht} can be derived starting from the invariant \cite{Bernabeu:1986fc},
\beq
I_1=\Tr [H^{\nu},H^E ]^3 = \Tr \left(    [H^{\nu}H^E -H^EH^{\nu}]^3                    \right)      
\label{I1}
\eeq
where the Hermitian matrices are defined as 
\beq
H^\nu = m^\nu m^{\nu \dag}, \ \ \ \ H^E = m^E m^{E \dag}
\eeq
In our conventions of section~\ref{conventions}, by explicit calculation one can verify the well known result that 
\beq
I_1 = -6i \Delta m_{\nu}^6 \Delta m_E^6 J
\label{I1c}
\eeq
where
\bea
\Delta m_{\nu}^6 & = & (m_3^2-m_1^2)(m_2^2-m_1^2)(m_3^2-m_2^2) =  m_3^2 m_2^2 \Delta m_{32}^2 \label{Dmnu} \\
\Delta m_E^6 & = & (m_{\tau}^2-m_e^2)(m_{\mu}^2-m_e^2)(m_{\tau}^2-m_{\mu}^2) \label{DmE} \\
J & = & \Im [U_{11}U_{22}U_{12}^*U_{21}^*]= s_{12}c_{12}s_{13}c_{13}^2s_{23}c_{23}\sin \delta \label{J}
\eea
The above results show that $I_1$ is basis invariant since it can be expressed in terms of 
physical masses and PMNS parameters. We are therefore free to evaluate $I_1$ in any basis.
For example, in the diagonal charged lepton mass basis,
one can shown that the quantity in Eq.\ref{I1} becomes,
\beq
I_1=\Tr [H^{\nu},H^E ]^3 = 6i \Delta m_E^6 \Im (H^{\nu *}_{12} H^{\nu }_{13} H^{\nu *}_{23} ) 
 \label{I1p}
\eeq
where in Eq.\ref{I1p}, the Hermitian matrix $H^\nu = m^\nu m^{\nu \dag}$ involves the neutrino
mass matrix $m^\nu$ in the basis where the charged lepton mass matrix is diagonal,
i.e. the basis of Eq.\ref{eq:mnu2}, where we find
\beq
 \Im (H^{\nu *}_{12} H^{\nu }_{13} H^{\nu *}_{23} ) 
= 24m_a^3m_b^3(n-1)\sin \eta
 \label{I1pp}
\eeq
From Eqs.\ref{I1c} and \ref{I1p} we find, after equating these two expressions,
\beq
J= -\frac{ \Im (H^{\nu *}_{12} H^{\nu }_{13} H^{\nu *}_{23} ) }{\Delta m_{\nu}^6}
= -\frac{ 24m_a^3m_b^3(n-1)\sin \eta }{m_3^2 m_2^2 \Delta m_{32}^2 }
\label{Jp}
\eeq
where we have used Eqs.\ref{Dmnu} and \ref{I1pp}.
From Eqs.\ref{J} and \ref{Jp} we find, after equating these two expressions, we find the exact relation
\beq
\sin \delta = -\frac{ 24m_a^3m_b^3(n-1)\sin \eta }{m_3^2 m_2^2 \Delta m_{32}^2 s_{12}c_{12}s_{13}c_{13}^2s_{23}c_{23}}
\label{sdelta}
\eeq
Note the minus sign in Eq.\ref{sdelta}, which means that, for $n>1$, the sign of $\sin \delta $
takes the opposite value to the sign of $\sin \eta$, in the convention we use to 
write our neutrino mass matrix in section~\ref{conventions}.
Since the denominator of Eq.\ref{sdelta} may be expressed in terms of input parameters,
using the exact results for the neutrino masses in Eqs.\ref{m3},\ref{m2} together with the exact results for the lepton mixing angles in Eqs.\ref{s13},\ref{t12},\ref{t23}, it is clear that Eq.\ref{sdelta} gives 
$\sin \delta $ in terms of input parameters in both the numerator and demominator.

In the SD approximation
in Eqs.\ref{m3SD}, \ref{m2SD} we find from Eq.\ref{Jp},
\beq
J\approx - \frac{1}{9}\frac{m_2}{m_3}(n-1)\sin \eta ,
\label{Jpp}
\eeq
where the minus sign in Eq.\ref{Jpp} again clearly shows the anti-sign correlation of $\sin \delta$ and 
$\sin \eta$, where $\eta $ is the input phase which appears in the neutrino mass matrix 
in Eq.\ref{eq:mnu2} and leptogenesis in Eq.\ref{BAU}. In other words the BAU is proportional to 
$-\sin \delta$ if the lightest right-handed neutrino is the one dominantly responsible for the
atmospheric neutrino mass $N_1 = N_{\rm atm}$ . In this case the observed matter Universe
requires $\sin \delta$ to be negative in order to generate a positive BAU.
It is interesting to note that, up to a negative factor, the sine of the leptogenesis phase $ \eta$
is equal to the sine of the oscillation phase $ \delta$, so the observation the CP violation 
in neutrino oscillations is directly responsible for the CP violation in the early Universe, in the LS model.

\subsection{The Majorana Invariant}

The Majorana invariant may be defined by \cite{Branco:1986gr},
\beq
I_2=\Im \Tr [H^Em^{\nu} m^{\nu \dagger}m^{\nu}H^{E*}m^{\nu \dagger}]  
\label{I2}
\eeq
In our conventions of section~\ref{conventions}, this may be written,
\beq
I_2 = \Im \Tr [U^{\dagger}H^E_{\rm diag}Um^{\nu 3 }_{\rm diag} U^T H^{E}_{\rm diag}U^* m^{\nu}_{\rm diag}]  
\label{I2b}
\eeq
By explicit calculation we find an exact but rather long expression which is basis invariant since it involves
physical masses and PMNS parameters. We do not show the result here since it is rather long and not very illuminating and also not so relevant since Majorana CP violation is not going to be measured
for a very long time.
However, since
$m_e^2\ll m_{\mu}^2\ll m_{\tau}^2$, we may neglect $m_e^2$ and $m_{\mu}^2$ compared to 
$m_{\tau}^2$, and also drop $s^2_{13}$ terms, to give the compact result,
\beq
I_2 \approx  -m^4_{\tau} m_2m_3\Delta m^2_{32}c^2_{13}c^2_{23}
[c^2_{12} s^2_{23} (\sin \beta) +2 c_{12}c_{23}s_{12}s_{13}s_{23}\sin (\beta + \delta)     ].
\label{I2c}
\eeq
Notice that $I_2$ is zero if both $\beta$ and $\delta$ are zero or $\pi$, so it is indeed sensitive to Majorana CP violation arising from $\beta$. Indeed $I_2$ is roughly proportional to $\sin \beta $ if the 
$s_{13}$ term is also neglected,
\beq
I_2 \approx  -m^4_{\tau} m_2m_3\Delta m^2_{32}c^2_{13}c^2_{23}
c^2_{12} s^2_{23} (\sin \beta) .
\label{I2d}
\eeq

We are free to evaluate $I_2$ in any basis.
For example, in the diagonal charged lepton mass basis,
the quantity in Eq.\ref{I2} becomes,
\beq
I_2=\Im \Tr [H^E_{\rm diag}m^{\nu} m^{\nu \dagger}m^{\nu}H^E_{\rm diag} m^{\nu \dagger}]  
\label{I2e}
\eeq
where in Eq.\ref{I2e}, the neutrino
mass matrix $m^\nu$ is in the basis where the charged lepton mass matrix is diagonal,
i.e. the basis of Eq.\ref{eq:mnu2}. Evaluating Eq.\ref{I2e} we find the exact result,
\beq
I_2
= m_a m_b \sin \eta \left[
-4m_a^2 (m^2_{\tau}-m^2_{\mu})^2
+m_b^2 \left(
m^2_{\tau}(2n+1)(n-2)
-m^2_{\mu}n(2n-5)
-2m_e^2(n-1)
\right)^2
\right]
 \label{I2f}
\eeq
If we neglect $m_e^2$ and $m_{\mu}^2$, Eq.\ref{I2f} becomes approximately,
\beq
I_2
\approx  m^4_{\tau} m_a m_b \sin \eta \left[
-4m_a^2 
+m_b^2 (2n+1)^2(n-2)^2
\right].
 \label{I2g}
\eeq
From Eqs.\ref{I2d} and \ref{I2g} we find, after equating these two expressions,
one finds an approximate formula for the sine of the Majorana phase, 
\beq
\sin \beta \approx  \frac{ m_a m_b [4m_a^2 -m_b^2 (2n+1)^2(n-2)^2]  \sin \eta}
{ m_2m_3\Delta m^2_{32}c^2_{13}c^2_{23}c^2_{12} s^2_{23}}
\label{sbeta}
\eeq
Since the denominator of Eq.\ref{sbeta} may be expressed in terms of input parameters,
using the exact results for the neutrino masses in Eqs.\ref{m3},\ref{m2} together with the exact results for the lepton mixing angles in Eqs.\ref{s13},\ref{t12},\ref{t23}, it is clear that Eq.\ref{sbeta} gives 
$\sin \beta $ in terms of input parameters in both the numerator and demominator.
For low values of $n$ (e.g. $n=3$) the sign of $\sin \beta$ is the same as the sign of $\sin \eta$
and hence the opposite of the sign of $\sin \delta$ given by Eq.\ref{sdelta}.

It is worth recalling at this point that our Majorana phases are 
in the convention of Eq.\ref{Maj}, namely $P={\rm diag}(e^{i\frac{\beta_1}{2}},e^{i\frac{\beta_2}{2}},1)$,
where we defined $\beta=\beta_2$ and $\beta_1$ is unphysical since $m_1=0$.
In another common convention the Majorana phases are by given by 
$P={\rm diag}(1,e^{i\frac{\alpha_{21}}{2}},e^{i\frac{\alpha_{31}}{2}})$, which are related to ours by
$\alpha_{21}=\beta_2-\beta_1$ and $\alpha_{31}=-\beta_1$. For the case at hand, where $m_1=0$, one finds 
$\beta  = \alpha_{21}-  \alpha_{31}$ 
to be the only Majorana phase having any physical significance (e.g. which enters the formula for neutrinoless double beta decay).
This is the phase given by Eq.\ref{sbeta}.

Eq.\ref{sbeta} is independent of $s_{13}$ since we have dropped those terms.
It is only therefore expected to be accurate to about 15\%, which is acceptable, given that 
the Majorana phase $\beta$ is practically impossible to
measure in the forseeable future for the case of a normal mass hierarchy with the lightest neutrino mass $m_1=0$.
However, if it becomes necessary in the future to have a more accurate result, this can be obtained by equating
Eq.\ref{I2c} with Eq.\ref{I2g}, which would yield an implicit formula for $\beta$ which is 
accurate to about 3\%:
\beq
\sin \beta  +2 c^{-1}_{12}c_{23}s_{12}s_{13}s^{-1}_{23}\sin (\beta + \delta) \approx  \frac{ m_a m_b [4m_a^2 -m_b^2 (2n+1)^2(n-2)^2]  \sin \eta}
{ m_2m_3\Delta m^2_{32}c^2_{13}c^2_{23}c^2_{12} s^2_{23}}.
\label{sbetap}
\eeq

\section{Exact Sum Rules}
\label{exactsumrules}

The formulas in the previous section give the observable physical neutrino masses and the PMNS
angles and phases in terms of fewer input parameters $m_a$, $m_b$, $n$ and $\eta$. In particular, the exact results for the neutrino masses are given in Eqs.\ref{m3},\ref{m2},\ref{m1},
the exact results for the lepton mixing angles are given in Eqs.\ref{s13},\ref{t12},\ref{t23}
and the exact result for the CP violating Dirac oscillation phase is given in Eq.\ref{sdelta},
while the Majorana phase is given approximately by Eq.\ref{sbeta}.
These 8 equations for the 8 observables cannot be inverted to give the 4 input parameters in terms of the 8 physical parameters since there are clearly fewer input parameters than observables. 
On the one hand, this is good, since it means that the littlest seesaw has 4 predictions, on the other hand
it does mean that we have to deal with a 4 dimensional input parameter space.
Later we shall impose additional theoretical considerations, which shall reduce this parameter space to just 
2 input parameters, yielding 6 predictions, but for now we consider the 4 input paramaters.
In any case it is not obvious that one can derive any sum rules where input parameters are eliminated, and only relations between physical observables remain. 

Nevertheless in this model there are such sum rules, i.e. relations between physical observables not involving the input parameters.
An example of such a sum rule is Eq.\ref{t12}, which we give below in three equivalent exact forms,
\beq
\tan \theta_{12} = \frac{1}{\sqrt{2}}\sqrt{1-3s^2_{13}}\ \ \ \ {\rm or} \ \ \ \ 
\sin \theta_{12}= \frac{1}{\sqrt{3}}\frac{\sqrt{1-3s^2_{13}}}{c_{13}} \ \ \ \ {\rm or} \ \ \ \ 
\cos \theta_{12}= \sqrt{\frac{2}{3}}\frac{1}{c_{13}}
\label{t12p}
\eeq
This sum rule is in fact common to all TM1 models, and is therefore also applicable to the LS models which 
predict TM1 mixing. Similarly TM1 predicts the so called atmospheric sum rule, also applicable to the LS models.
This arises from the fact that the first column of the PMNS matrix is the same as the first column of the TB matrix in Eq.\ref{TB}. Indeed, by comparing the magnitudes of the elements in the first column of Eq.\ref{eq:Umatrix} to those in the first column of Eq.\ref{eq:UmatrixPDG}, we obtain,
\bea
|U_{e1}| &=& c_{12} c_{13} = \sqrt{\frac{2}{3}} \label{Ue1}\\
|U_{\mu 1}| &=& | - s_{12} c_{23} - c_{12} s_{13} s_{23} e^{i\delta}|= \sqrt{\frac{1}{6}}  \label{Umu1}\\
|U_{\tau 1}| &=& |s_{12} s_{23} - c_{12} s_{13} c_{23} e^{i\delta}| = \sqrt{\frac{1}{6}}  \label{Utau1}
\eea
Eq.\ref{Ue1} is equivalent to Eq.\ref{t12p} while 
Eqs.\ref{Umu1} and \ref{Utau1} lead to equivalent 
mixing sum rules which can be expressed as an exact relation for $\cos \delta$ in terms of the other lepton mixing angles
\cite{Albright:2008rp},
\beq
\cos \delta = - \frac{\cot 2\theta_{23}(1-5s^2_{13})}{2\sqrt{2}s_{13}\sqrt{1-3s^2_{13}}}
\label{TM1sum}
\eeq
Note that, for maximal atmospheric mixing, $\theta_{23}=\pi/4$, we see that 
$\cot 2\theta_{23}=0$ and therefore 
this sum rule predicts $\cos \delta =0$,
corresponding to maximal CP violation $\delta = \pm \pi/2$.
The prospects for testing the TM1 atmospheric sum rules Eqs.\ref{t12p},\ref{TM1sum}
in future neutrino facilities 
was discussed in \cite{Ballett:2013wya}.

The LS model also the predicts additional sum rules beyond the TM1 sum rules
that arise from the structure of the Dirac mass matrix in Eq.\ref{mDn}.
Recalling that the PMNS matrix is written in Eq.\ref{DM} as $U=VP$, where $V$ is the 
the CKM-like part and $P$ contains the Majorana phase $\beta$,
the LS sum rules are \cite{King:2013iva},
\bea
\frac{|V_{\mu 3}V_{e2}-V_{\mu 2}V_{e3}|}{|V_{\tau 3}V_{e2}-V_{\tau 2}V_{e3}|}&=&1 \label{LS1}\\
\frac{|e^{i\beta}m_2V_{\mu 2}V_{e2}+m_3V_{\mu 3}V_{e3}|}
{|e^{i\beta}m_2V^2_{e2}+m_3V^2_{e3}|}&=&n \label{LSn}\\
\frac{|e^{i\beta}m_2V_{\tau 2}V_{e2}+m_3V_{\tau 3}V_{e3}|}
{|e^{i\beta}m_2V^2_{e2}+m_3V^2_{e3}|} &=&n-2 \label{LSn-2}
\eea
where the sum rule in Eq.\ref{LS1} is independent of both $n$ and $\beta$.
We emphasise again that the matrix elements $V_{\alpha i}$ refer to the first matrix 
on the right-hand side of Eq.\ref{eq:UmatrixPDG} (i.e. without the Majorana matrix).
Of course similar relations apply with $U$ replacing $V$ everywhere
and the Majorana phase $\beta$ disappearing, being absorbed into the PMNS matrix $U$, but we prefer to exhibit the Majorana phase dependence explicitly.
However the LS sum rule in Eq.\ref{LS1} is equivalent to the TM1 sum rule in Eq.\ref{TM1sum},
as seen by explicit calculation. The other LS sum rules in Eqs.\ref{LSn} and \ref{LSn-2} involve the phase $\beta$
and are not so interesting.

\section{The reactor and atmospheric angles}
\label{ra}
Since the solar angle is expected to be very close to its tribimaximal value,
according to the TM1 sum rules in Eq.\ref{t12p}, independent
of the input parameters, in this section we focus on the analytic predictions for the reactor and atmospheric angles,
starting with 
the accurately measured reactor angle which is very important for pinning down the input parameters of the LS model.

\subsection{The reactor angle}
The exact expression for the reactor angle in Eq.\ref{s13} is summarised below,
\beq
\sin \theta_{13}  =  \frac{1}{\sqrt{6}}\left(1-\sqrt{\frac{1 }{1+t^2}}       \right)^{1/2},
\label{s13p}
\eeq
where from Eqs.~\ref{xyz},\ref{t},\ref{B},\ref{A},
\beq
t =   \frac{2 \sqrt{6}m_b  (n-1)}{2| m_a+ m_b e^{i\eta} (n-1)^2|\cos (A - B) -3m_b  \cos B}  
\label{tp}
\eeq
where 
\beq
\tan B =   \frac{ 2| m_a+ m_b e^{i\eta} (n-1)^2|\sin A}{3m_b+2| m_a+ m_b e^{i\eta} (n-1)^2|\cos A}  
\label{Bp}
\eeq
and 
\beq
A= \arg [m_a+ m_b e^{i\eta} (n-1)^2 ] - \eta .\label{Ap}
\eeq
The above results are exact and necessary for precise analysis of the model, especially for large $n$
(where $n$ is in general a real and continuous number).
We now proceed to derive some approxinate formulae which can give useful insight.

The SD approximations in Eqs.\ref{m3SD},\ref{m2SD} show 
that $m_b/m_a\approx (2/3)m_2/m_3$. This suggests that we can make an expansion in
$m_b/m_a$, or simply drop $m_b$ compared to $m_a$, as a leading order approximation,
which implies $\tan B \approx \tan A $ and hence $\cos (A-B)\approx 1$.
Thus Eq.\ref{tp} becomes,
\beq
t  \approx \sqrt{6}  \frac{m_b(n-1)}{| m_a+ m_b e^{i\eta} (n-1)^2|} 
\label{tpp}
\eeq
where we have kept the term proportional to $m_b (n-1)^2$, since the smallness of $m_b$
may be compensated by the factor $(n-1)^2$ for $n>1$.
Eq.\ref{tpp} shows that $t\ll 1$, hence we may expand Eq.\ref{s13p} to leading order in $t$, 
\beq
\sin \theta_{13}  \approx  \frac{t}{2\sqrt{3}}+{\cal O}(t^3).
\label{s13pp}
\eeq
Hence combining Eqs.\ref{tpp} and \ref{s13pp}, we arrive at our approximate form for the sine of the reactor
angle,
\beq
\sin \theta_{13}  \approx  \frac{1}{\sqrt{2}}  \frac{m_b(n-1)}{| m_a+ m_b e^{i\eta} (n-1)^2|} .
\label{s13ppp}
\eeq
For low values of $(n-1)$ such that $m_b(n-1)^2\ll m_a$, Eq.\ref{s13ppp} simplifies to,
\beq
\sin \theta_{13} \approx (n-1) \frac{\sqrt{2}}{3}  \frac{m_2}{m_3}
\label{s13ppppp}
\eeq
using the SD approximations in Eqs.\ref{m3SD},\ref{m2SD} 
that $m_b/m_a\approx (2/3)m_2/m_3$, valid to 10\% accuracy.
For example, the result shows that for the original CSD \cite{King:2005bj}, 
where $n=1$, implies $\sin \theta_{13}=0$,
while for CSD(2) \cite{Antusch:2011ic} 
(i.e. $n=2$) we have $\sin \theta_{13} \approx \frac{\sqrt{2}}{3}  \frac{m_2}{m_3}$, 
leading to $\theta_{13}\approx 4.7^{\circ}$ which is too small.
For  CSD(3) \cite{King:2013iva} (i.e. $n=3$) we have $\sin \theta_{13} \approx \frac{2\sqrt{2}}{3}  \frac{m_2}{m_3}$, 
leading to $\theta_{13}\approx 9.5^{\circ}$, in rough agreement with the observed value of $\theta_{13}\approx 8.5^{\circ}$, within the accuracy of our approximations.
We conclude that these results show how $\sin \theta_{13}\sim {\cal O}(m_2/m_3)$
can be achieved, with values increasing with $n$, and 
confirm that $n\approx 3$ gives the best fit to the reactor angle.
We emphasise that the approximate formula in Eq.\ref{s13ppppp} has not been 
written down before, and that the exact results in Eqs.\ref{s13p},\ref{tp},\ref{Bp},\ref{Ap} 
are also new and in perfect agreement with the numerical results in Table~\ref{tab:model}.

\subsection{The atmospheric angle}
The exact expression for the atmospheric angle in Eq.\ref{t23} is summarised below,
\beq
\tan \theta_{23} =\frac{|1+\epsilon^{\nu}_{23}|}{|1-\epsilon^{\nu}_{23}|}
\label{t23p}
\eeq
where 
\beq
\epsilon^{\nu}_{23} = \sqrt{\frac{2}{3}}t^{-1} 
\left[\sqrt{1+t^2 } -1 \right] e^{-iB}
\label{epsnup}
\eeq
and $t$ and $B$ were summarised in Eqs.\ref{tp},\ref{Bp},\ref{Ap}.
The above results, which are exact, show that 
the atmospheric angle is maximal for $B \approx \pm \pi/2$,
as noted previously.

We may expand Eq.\ref{epsnup} to leading order in $t$, 
\beq
\epsilon^{\nu}_{23} \approx \frac{t}{\sqrt{6}}e^{-iB} +{\cal O}(t^3).
\label{epsnupp}
\eeq
Hence combining Eqs.\ref{t23p},\ref{epsnupp},
we arrive at an approximate form for the tangent of the atmospheric angle,
\beq
\tan \theta_{23} \approx \frac{|1+ \frac{t}{\sqrt{6}}e^{-iB}|}{|1- \frac{t}{\sqrt{6}}e^{-iB}|}
\approx 1+2 \frac{t}{\sqrt{6}} \cos B,
\label{t23pp}
\eeq
where $t$ was approximated in Eq.\ref{tpp}.
We observed earlier that, for $m_b\ll m_a$,
$\tan B \approx \tan A $ and hence $A\approx B$.
Unfortunately it is not easy to obtain a reliable approximation for $A$ in Eq.\ref{Ap},
unless $n\simgt 1$ in which case $A\approx -\eta$.
However, for $n-1$ significantly larger than unity, this is not a good approximation.
For example for $n=3$ from Eq.\ref{Ap} we have,
\beq
A= \arg [m_a+ 4 m_b e^{i\eta} ] - \eta 
\label{App}
\eeq
Taking $m_b/m_a = 1/10$, this gives
\beq
A= \arg [1+ 0.4 e^{i\eta} ] - \eta 
\label{Appp}
\eeq
which shows that $A\approx -\eta$ is not a good approximation even though $m_b\ll m_a$.
If we set, for example, $\eta = 2\pi/3$, as in Table~\ref{tab:model}, then Eq.\ref{Appp} gives 
\beq
A=  0.41 - 2\pi/3  \approx - 0.53 \pi
\label{Apppp}
\eeq
which happens to be close to $-\pi/2$. Hence, since $A\approx B$, this choice of parameters 
implies $\cos B \approx 0$,
leading to approximately maximal atmospheric mixing from Eq.\ref{t23pp},
as observed in Table~\ref{tab:model}.
At this point it is also worth recalling that for maximal atmospheric mixing,
the TM1 sum rule in Eq.\ref{TM1sum} predicts that the cosine of the CP phase $\delta$
to be zero, corresponding to maximal Dirac CP violation $\delta = \pm \pi/2$,
as approximately found in Table~\ref{tab:model}.

\section{CSD(3) vacuum alignments from $S_4$}
\label{S4vac}
We saw from the discussion of the reactor angle, and in Table~\ref{tab:model},
that the solar alignment in Eq.\ref{Phias} for the particular choice $n=3$ was favoured.  
In this section we show how the desired alignments for $n=3$ can emerge from $S_4$
due to residual symmetries. Although the charged lepton alignments we discuss 
were also obtained previously from $A_4$
 \cite{King:2013iva}, the neutrino alignments  in Eq.\ref{Phias} for $n=3$
were not previously obtained from residual symmetries, and indeed we will see that 
they will arise from group elements which appear in $S_4$ but not $A_4$.

We first summarise the vacuum alignments that we desire:
\begin{equation}
\label{Phias2} 
\vev{\phi_{\rm atm}}
=v_{\rm atm} \begin{pmatrix}0 \\ 1 \\ 1\end{pmatrix},
 \qquad
\vev{\phi_{\rm sol}}
=v_{\rm sol}
\begin{pmatrix}1 \\ 3 \\ 1\end{pmatrix},
\end{equation}
in the neutrino sector as in Eq.\ref{Phias} with $n=3$, and,
\be
\langle \varphi_e \rangle =v_e 
\begin{pmatrix} 1\\0\\0 \end{pmatrix}  \ , \qquad
\langle \varphi_\mu \rangle =v_\mu 
\begin{pmatrix} 0\\1\\0 \end{pmatrix} \ , \qquad
\langle \varphi_\tau \rangle = v_\tau 
\begin{pmatrix} 0\\0\\1 \end{pmatrix} 
 \ .\label{Phil}
\ee
in the charged lepton sector.

For comparison we also give the tribimaximal alignments in Eq.\ref{Phi123}:
\begin{equation}
\label{Ph123p} 
\vev{\phi_1}
=v_1\begin{pmatrix}2 \\ -1 \\ 1\end{pmatrix},
 \qquad
 \vev{\phi_2}
=v_2\begin{pmatrix}1 \\ 1 \\ -1\end{pmatrix},
 \qquad
 \vev{\phi_3}
=v_3\begin{pmatrix}0 \\ 1 \\ 1\end{pmatrix}.
\end{equation}

We first observe that the charged lepton and the tribimaximal alignments individually preserve some
remnant symmetry of $S_4$, whose triplet representations are displayed explicitly in Appendix~\ref{S4}.
If we regard $\varphi_e$, $\varphi_{\mu}$, $ \varphi_{\tau}$ as each being a triplet $\bf{3}$
of $S_4$, then 
they each correspond to a different symmetry conserving direction of $S_4$, with,
\beq
a_2 \langle \varphi_e \rangle =  \langle \varphi_e \rangle, \ \ 
a_3 \langle \varphi_{\mu} \rangle = \langle \varphi_{\mu} \rangle, \ \ 
a_4 \langle \varphi_{\tau} \rangle = \langle \varphi_{\tau} \rangle.
\label{PhilS4}
\eeq
One may question the use of different residual symmetry generators of  $S_4$ to enforce
the different charged lepton vacuum alignments. However, this is analagous to what is usually assumed
in the direct model building approach when one says that the charged lepton sector preserves
one residual symmetry, while the neutrino sector preserves another residual symmetry.
In the direct case, it is clear that the lepton Lagrangian as a whole completely breaks the family symmetry,
even though the charged lepton and neutrino sectors preserve different residual symmetries.
In the indirect case here, we are taking this argument one step further, by saying that 
the electron, muon and tau sectors preserve different residual symmetries, while the charged lepton
Lagrangian as a whole completely breaks the family symmetry. However the principle is the same 
as in the direct models, namely that different sectors of the Lagrangian preserve different residual subgroups
of the family symmetry.

\begin{figure}[htb]
\centering
\includegraphics[width=0.6\textwidth]{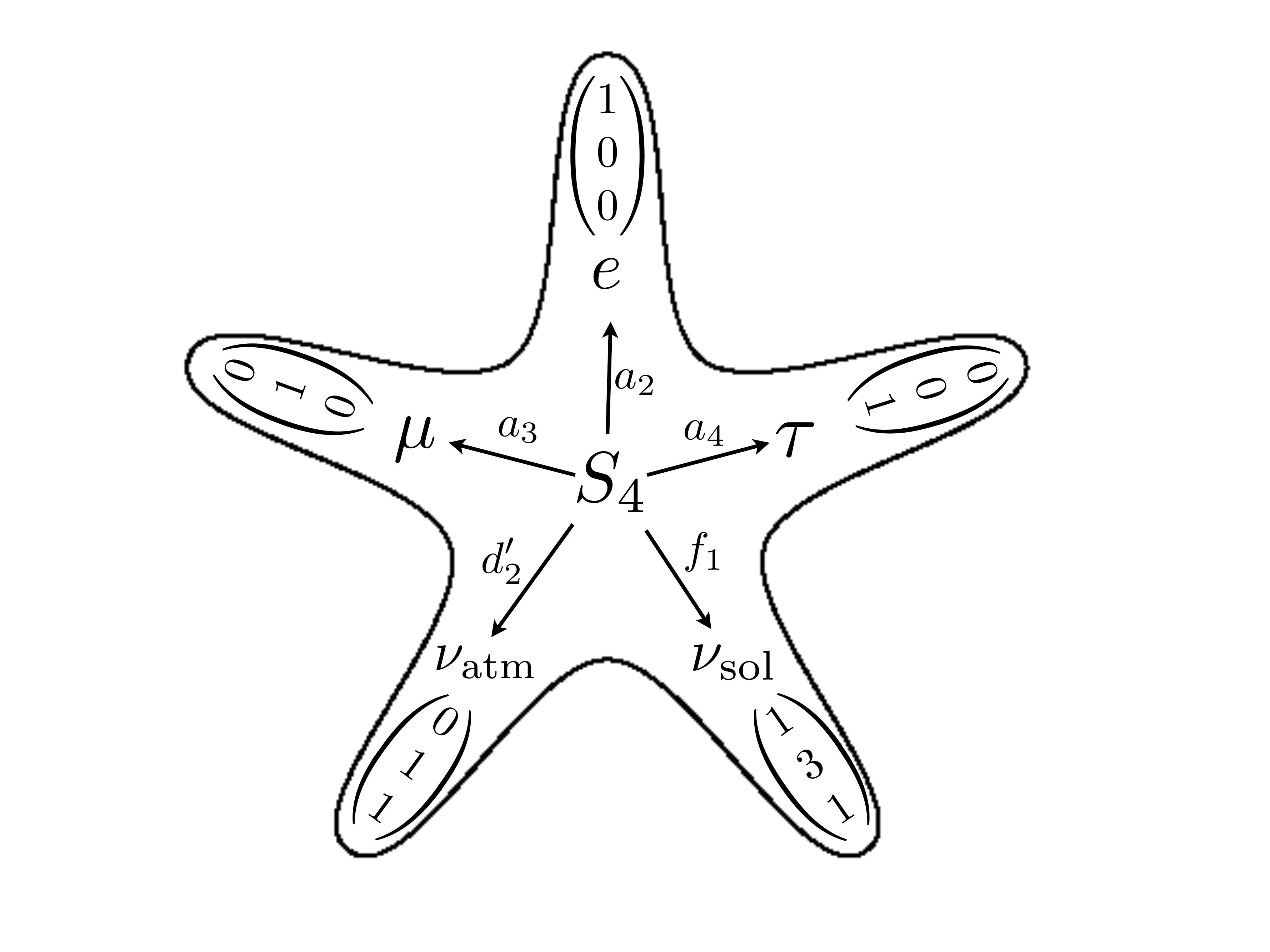}
\caption{\label{starfish}\small{
This starfish diagram summarises the residual $S_4$ symmetries respected by the vacuum alignments associated with the various physical sectors of the model. The residual symmetries are given by the preserved group elements 
$a_i,d'_2,f_1$ defined in Appendix~\ref{S4}.
}}
\end{figure}

The tribimaximal alignment $ \vev{\phi_2}$ is enforced by a combination of $d_2$ and $f_1$ being conserved,
\beq
d_2 \vev{\phi_2}=  \vev{\phi_2}, \ \ 
f_1 \vev{\phi_2}=  \vev{\phi_2},
\label{Phi2S4}
\eeq
which suggests that $\phi_2$ should be also identified as a triplet $\bf{3}$
of $S_4$. On the other hand, the tribimaximal alignment $ \vev{\phi_3}$ (which is the same as the atmospheric
alignment $\vev{\phi_{\rm atm}}$) may be enforced by symmetry if $\phi_3$ 
(i.e. $\phi_{\rm atm}$) is in the $\bf{3'}$ representation,
since then we see that,
\beq
d'_2 \vev{\phi_3}=  \vev{\phi_3},\ \ {\rm or} \ \ 
d'_2 \vev{\phi_{\rm atm}}=  \vev{\phi_{\rm atm}},
\label{Phi3S4}
\eeq
As in the case of the charged lepton sector, we see that different parts of the neutrino sector will preserve
different residual subgroups of the family symmetry $S_4$ for the tribimaximal alignments
$ \vev{\phi_2}$ and $\vev{\phi_3}=\vev{\phi_{\rm atm}}$. 

In order to obtain the alignments $\vev{\phi_1}$ and $\vev{\phi_{\rm sol}}$ we must depart from
the idea of residual symmetries and resort to dynamical terms in the potential that enforce orthogonality,
as discussed in section~\ref{indirect}.
However, once the tribimaximal alignments $\vev{\phi_2}$ and $\vev{\phi_3}$ have been accomplished,
the remaining tribimaximal alignment $\vev{\phi_1}$ is simple to obtain, see Fig.~\ref{perp}.
Similarly the general solar alignment in Eq.\ref{Phias} then follows from the orthogonality 
to $\vev{\phi_1}$, as is also clear from Fig.~\ref{perp}. 

We now observe that the particular solar alignment
$\vev{\phi_{\rm sol}}$ in Eq.\ref{Phias2} can be natually enforced by a symmetry argument
if $\phi_{\rm sol}$ is a triplet $\bf{3}$
of $S_4$ since then,
\beq
f_1 \vev{\phi_{\rm sol}}=   \vev{\phi_{\rm sol}},
\label{PhisolS4}
\eeq
which by itself constrains the alignment to be
$(1,m,1)$, for continuous real $m$. 
However orthogonality to $\vev{\phi_1}$ further constrains the alignment to be
$(1,n,n-2)$, for continuous real $n$.
Taken together, the constrained forms $(1,n,n-2)$ and $(1,m,1)$, 
fix $n=m$ and $n-2=1$, and hence $n=m=3$, corresponding to  
the alignment $(1,3,1)$ as desired in Eq.\ref{Phias2}.

To summarise we see that the desired alignments in Eqs.\ref{Phias2}
and \ref{Phil} emerge naturally from the residual symmetries of $S_4$, together with the simple
orthogonality conditions which can be 
readily obtained in models as in Eq.\ref{orthog}.
The residual $S_4$ symmetries involved in the various physical sectors of the model are summarised 
by the starfish shaped diagram in Fig.\ref{starfish}.

\section{A benchmark model with $S_4\times Z_3\times Z_3'$}
\label{benchmarkmodel}

\begin{table}
	\centering
$$
\begin{array}{||c||cccccccc||c||ccccccc||}
\hline \hline
&\theta &e^c&\mu^c & \tau^c
&\varphi_e&\varphi_\mu&\varphi_\tau
&H_{1}&L&H_{2}&\phi_{\rm atm}&\phi_{\rm sol}
&\nu^c_{\rm atm}&\nu^c_{\rm sol}
&\xi_{\rm atm}&\xi_{\rm sol} \\  \hline
\hline
S_4 & {\bf 1} & {\bf 1} & {\bf 1} & {\bf 1} & {\bf 3} & {\bf 3} & {\bf 3}  & {\bf 1}
&{\bf 3} & {\bf 1} & {\bf 3'} & {\bf 3} & {\bf 1'}& {\bf 1}& {\bf 1}& {\bf 1} 
\\[2mm] \hline  \hline
Z_3 & 1 & \omega  & \omega^2 & 1&\omega^2&\omega &1&1 
&1& 1 &\omega &\omega^2&\omega^2 & \omega  &\omega^2 &\omega \\[2mm] \hline 
Z'_3 & \omega &  \omega^2 & \omega &1 & \omega^2  &\omega&1&1&1&1 &\omega^2&1& \omega 
&1&\omega&1\\[2mm] \hline
R&0&1&1&1&0&0&0&0&1&0&0&0&1&1&0&0 \\ 
\hline
\end{array}
$$
\caption{\label{tab-S4}Lepton, Higgs and flavon superfields 
and how they may transform under the symmetries relevant for the Yukawa sector of the model.
The $U(1)_R$ symmetry,
under which all the leptons have a charge of unity while the Higgs and flavons have zero charge,
are also shown in the Table. }
\end{table}

We now present a model based on $S_4\times Z_3\times Z'_3$, 
which can reproduce the numerical benchmark discussed in section~\ref{benchmark}.
The $S_4$ will help produce the vacuum alignments with $n=3$, as discussed in the previous section,
the $Z_3$ will help to fix $\eta = 2\pi/3$ while the $Z'_3$ will be responsible for the charged lepton
mass hierarchy.
This will yield the most predictive and successful version of the LS model, corresponding to the numerical
results in Table~\ref{tab:model}, perfectly reproduced by the exact analytic results,
where the two remaining free parameters $m_a$
and $m_b$ are used to fix the neutrino mass squared differences. The entire PMNS matrix then emerges
as a parameter free prediction, corresponding to the CSD(3) benchmark discussed in 
section~\ref{benchmark}.

With the alignments in Eqs.\ref{Phias2}
and \ref{Phil}, arising as a consequence of $S_4$ 
residual symmetry, summarised by the starfish diagram in Fig.\ref{starfish}, together with 
simple orthogonality conditions, as further discussed in the next section,
we may write down the superpotential of the starfish lepton model, as a supersymmetric version
\footnote{It is trivial to convert this superpotential to a non-supersymmetric
Lagrangian by interpreting the leptons as fermions and the Higgs and flavons as scalars,
rather than superfields. In the non-supersymmetric case it is possible to identify
$H_2$ as the CP conjugate of the Higgs doublet $H_1$.
It is also possible to absorb this Higgs doublet into the flavon fields so that 
$\phi_i$ and $\varphi_{\alpha}$ represent five $S_4$ triplets of Higgs doublets,
corresponding to 15 Higgs doublets, in which case one power of $\Lambda$ would be removed from 
the denominator each term.} 
of the LS Lagrangian in Eq.\ref{LS}, 
\beq
{W_{S_4}^{\rm yuk}}=\frac{1}{\Lambda}H_2({L}.\phi_{\rm atm})\nu^c_{\rm atm}
+\frac{1}{\Lambda}H_2({L}.\phi_{\rm sol})\nu^c_{\rm sol} 
+\xi_{\rm atm}\nu^c_{\rm atm}\nu^c_{\rm atm}
 +\xi_{\rm sol}\nu^c_{\rm sol}\nu^c_{\rm sol}
\label{LS3}
\eeq
\begin{equation}
+\frac{1}{\Lambda}H_1 ({L}.\varphi_{\tau})\tau^c
+ \frac{1}{\Lambda^2}  \theta H_1 ({L}.\varphi_{\mu}) \mu^c
+\frac{1}{\Lambda^3}  \theta^2 H_1( {L}.\varphi_e) e^c
\label{l}
\end{equation}
where only these terms are allowed by the charges in Table~\ref{tab-S4},
where $L$ are the three families of electroweak lepton doublets unified into a single triplet of $S_4$,
$\nu^c_{\rm atm}$ and $\nu^c_{\rm sol}$ are the (CP conjugated) right-handed neutrinos which are
singlets of $S_4$, as are the two Higgs doublets $H_{1,2}$,
while $\phi_i$ and $\varphi_{\alpha}$ are $S_4$ triplet scalar flavons
with the vacuum alignments in Eqs.\ref{Phias2} and \ref{Phil}.
Note that, according to the arguments of the previous
section, all flavons are in the $\bf{3}$ representation
apart from the atmospheric flavon which must be in the $\bf{3'}$ which implies that the atmospheric right-handed
neutrino must be in the $\bf{1'}$ of $S_4$. 
We have introduced Majoron singlets $\xi_i$ whose VEVs will generate the diagonal right-handed neutrino masses.
We have suppressed all dimensionless Yukawa coupling constants, assumed to be of order unity,
with the charged lepton mass hierarchy originating from powers of 
$\theta$ which is an $S_4$ singlet. 
The corresponding powers of the mass scale $\Lambda$ keeps track of the mass dimension of each term.

Inserting the vacuum alignments in Eqs.\ref{Phias2}
and \ref{Phil} into Eqs.\ref{LS3} and \ref{l}, we obtain the neutrino and charged lepton Yukawa matrices,
with the rigid CSD(3) structure, 
\begin{equation}
	Y^{\nu} =\pmatr{0 & y_{\rm sol} \\ y_{\rm atm} & 3y_{\rm sol} \\ y_{\rm atm} & y_{\rm sol} },	\ \ \ \ 
	Y^{E} =\pmatr{y_e & 0 & 0 \\ 0 & y_{\mu} & 0 \\ 0 & 0 & y_{\tau}},
	\label{Y}
\end{equation}
where 
\beq
y_{\rm atm}\sim \frac{v_{\rm atm}}{\Lambda}, \ \ y_{\rm sol}\sim \frac{v_{\rm sol}}{\Lambda}, \ \ 
y_{\tau}\sim \frac{v_{\tau}}{\Lambda}, \ \ y_{\mu}\sim \frac{v_{\mu}\vev{\theta}}{\Lambda^2}, \ \ 
y_{e}\sim \frac{v_{e}\vev{\theta}^2}{\Lambda^3}.
\eeq
Note that we have a qualitative understanding of the charged lepton mass hierarchy as being due to successive powers of $\theta$, but there is no predictive power (for charged lepton masses) due to the arbitrary flavon VEVs and undetermined order unity dimensionless Yukawa couplings which we suppress.

When the Majorons get VEVs, the last two terms in Eq.\ref{LS3} will lead to the diagonal heavy Majorana mass matrix,
\beq
M_{R}=
\left( \begin{array}{cc}
M_{\rm atm} & 0 \\
0 & M_{\rm sol}
\end{array}
\right),
\label{MR3}
\eeq
where $M_{\rm atm}\sim \vev{\xi_{\rm atm}}$ and $M_{\rm sol}\sim \vev{\xi_{\rm sol}}$.
With the above seesaw matrices, we now have all the ingredients to reproduce the CSD(3) benchmark
neutrino mass matrix in Eq.~\ref{eq:mnu2p}, apart from the origin of the phase $\eta=2\pi/3$, which 
arises from vacuum alignment as we discuss in the next section.

\section{Vacuum alignment in the $S_4\times Z_3\times Z'_3$ model}
\label{benchmarkmodelvacuum}

We have argued in section~\ref{S4vac} that in general the vacuum the alignments in Eqs.\ref{Phias2}
and \ref{Phil}, arise as a consequence of $S_4$ 
residual symmetry, summarised by the starfish diagram in Fig.\ref{starfish}, together with 
simple orthogonality conditions. It remains to show how this can be accomplished, together with the Majoron VEVs,
by explicit superpotential alignment terms.

The charged lepton alignments in Eq.\ref{Phil} which naturally arise as a consequence of $S_4$,
can be generated from the simple terms,
\be
W_{S_4}^{\mathrm{flav},\ell}=
A_e \varphi_e \varphi_e + A_\mu \varphi_\mu \varphi_\mu + A_\tau \varphi_\tau \varphi_\tau.  
\label{a4-align-charged}
\ee
where $A_{l}$ are $S_4$ triplet $\bf{3}$ driving fields with necessary $ Z_3\times Z'_3$ charges to 
absorb the charges of $\phi_l$ so as to allow the terms in Eq.\ref{a4-align-charged}.
F-flatness then leads to the desired charged lepton flavon alignments in Eq.\ref{Phil}
due to,
\be
 \begin{pmatrix}
\langle \phi_{l} \rangle_2  \langle \phi_{l} \rangle_3 \\
\langle \phi_{l} \rangle_3  \langle \phi_{l} \rangle_1 \\
\langle \phi_{l} \rangle_1  \langle \phi_{l} \rangle_2 
\end{pmatrix}
 ~=~ 
\begin{pmatrix} 0\\0\\0
\end{pmatrix}
\ee
for $l=e,\mu,\tau$.

The vacuum alignment of the neutrino flavons involves the additional tribimaximal flavons $\phi_i$
with the orthogonality
terms in Eq.\ref{orthog}, 
\beq
W_{S_4}^{\mathrm{flav,perp}}=
 O_{ij}\phi_i \phi_j + O_{\rm sol}\phi_{\rm sol} \phi_1
\label{orthogp}
\eeq
where we desire the tribimaximal alignments in Eq.\ref{Ph123p} 
and as usual we identify $\phi_{\rm atm}\equiv \phi_3$.
We shall assume CP conservation with all triplet flavons acquiring real CP conserving VEVs.
Since there is some freedom in the choice of $\phi_{1,2}$ 
charges under $ Z_3\times Z'_3$, we leave them unspecified.
The singlet driving fields $O_{ij}$ and $O_{\rm sol}$ have $R=2$ and 
$Z_3\times Z'_3$ charges fixed by the (unspecified) $\phi_i$ charges,

The tribimaximal alignment for $\phi_2$ in the $\bf{3}$ in Eq.\ref{Ph123p} naturally arises as a consequence of $S_4$
from the simple terms,
\beq
W_{S_4}^{\mathrm{flav,TB2}}= 
A_{2} (g_2 \phi_{2}\phi_{2}  
+ g_2' \phi_{2} \xi_{2}  ).
\label{s4-flavon-nu2}
\eeq
where $A_2$ is an $R=2$, $S_4$ triplet $\bf{3}$ driving field and $\xi_2$ is a singlet,
with the same (unspecified) $ Z_3\times Z'_3$ charge as $\phi_2$.
F-flatness leads to, 
\be
2 g_2 \begin{pmatrix}
\langle \phi_{2} \rangle_2  \langle \phi_{2} \rangle_3 \\
\langle \phi_{2} \rangle_3  \langle \phi_{2} \rangle_1 \\
\langle \phi_{2} \rangle_1  \langle \phi_{2} \rangle_2 
\end{pmatrix}
+ g_2' \langle \xi_{2}  \rangle 
\begin{pmatrix}
\langle \phi_{2} \rangle_1\\
\langle \phi_{2} \rangle_2\\
\langle \phi_{2} \rangle_3
\end{pmatrix} ~=~ 
\begin{pmatrix} 0\\0\\0
\end{pmatrix},
\label{s4-flavon2}
\ee
leading to the tribimaximal alignment for $\phi_2$ in Eq.\ref{Ph123p}.
Note that in general the alignment derived from these $F$-term conditions is 
$\langle  \phi_{2} \rangle \propto (\pm1 ,\pm 1,\pm1)^T$. These are all equivalent.
For example $(1,1,-1)$ is related to permutations of the minus sign by $S_4$ transformations. The other choices can be obtained from these by simply multiplying an overall phase which would also change the sign of the $\xi_{2}$ VEV.

The tribimaximal alignment for $\phi_{\rm atm}\equiv \phi_3$ in the $\bf{3'}$
in Eq.\ref{Ph123p} naturally arises from
\beq
W_{S_4}^{\mathrm{flav,atm}}= 
A_{3} (g_3 \phi_{3}\phi_{3}  
+ g_3' \phi_{e} \xi_{3}  ),
\label{s4-flavon3}
\eeq
where $A_3$ is an $S_4$ triplet $\bf{3}$ driving field and $\xi_3$ is a singlet,
with suitable $ Z_3\times Z_3^{\theta}$ charges assigned to all the fields so as to allow only these terms.
F-flatness leads to, 
\be
2 g_3 \begin{pmatrix}
\langle \phi_{3} \rangle_2  \langle \phi_{3} \rangle_3 \\
\langle \phi_{3} \rangle_3  \langle \phi_{3} \rangle_1 \\
\langle \phi_{3} \rangle_1  \langle \phi_{3} \rangle_2 
\end{pmatrix}
+ g_3' \langle \xi_{2}  \rangle 
\begin{pmatrix}
\langle \phi_{e} \rangle_1\\
\langle \phi_{e} \rangle_2\\
\langle \phi_{e} \rangle_3
\end{pmatrix} ~=~ 
\begin{pmatrix} 0\\0\\0
\end{pmatrix},
\label{s4-flavon2}
\ee
which, using the orthogonality of $\phi_2$ and $\phi_3$ using Eq.\ref{orthogp}
and the pre-aligned electron flavon in Eq.\ref{Phil},
leads to the tribimaximal alignment for $\phi_{\rm atm}\equiv \phi_3$ in Eq.\ref{Ph123p}.
The tribimaximal alignment for $\phi_1$ then follows directly from the orthogonality 
conditions resulting from Eq.\ref{orthogp}. 

The solar flavon alignment comes from the terms,
\beq
W_{S_4}^{\mathrm{flav,sol}}= 
A_{sol} (g_{\rm sol} \phi_{\rm sol}\phi_{\rm sol}  
+ g_{\rm sol}' \phi_{\rm sol} \xi'_{\rm sol} + g_{\mu} \phi_{\mu} \xi_{\mu}).
\label{s4-flavon-nusol}
\eeq
F-flatness leads to, 
\be
2 g_{\rm sol} \begin{pmatrix}
\langle \phi_{\rm sol} \rangle_2  \langle \phi_{\rm sol} \rangle_3 \\
\langle \phi_{\rm sol} \rangle_3  \langle \phi_{\rm sol} \rangle_1 \\
\langle \phi_{\rm sol} \rangle_1  \langle \phi_{\rm sol} \rangle_2 
\end{pmatrix}
+ g_{\rm sol}' \langle \xi'_{\rm sol}  \rangle 
\begin{pmatrix}
\langle \phi_{\rm sol} \rangle_1\\
\langle \phi_{\rm sol} \rangle_2\\
\langle \phi_{\rm sol} \rangle_3
\end{pmatrix}
+ g_{\rm \mu} \langle \xi_{\mu}  \rangle 
\begin{pmatrix}
\langle \phi_{\mu} \rangle_1\\
\langle \phi_{\mu} \rangle_2\\
\langle \phi_{\mu} \rangle_3
\end{pmatrix}
 ~=~ 
\begin{pmatrix} 0\\0\\0
\end{pmatrix},
\label{s4-flavonsol}
\ee

which, using the pre-aligned muon flavon in Eq.\ref{Phil},
leads to the form $(1,m,1)$ for $\vev{\phi_{\rm sol}}$,
with $m$ unspecified, depending on the muon flavon VEV.
On the other hand the last term in Eq.\ref{orthogp} gives the general 
CSD(n) form in Eq.\ref{Phias}, $(1,n,n-2)$ for $\vev{\phi_{\rm sol}}$.
The two constrained forms $(1,n,n-2)$ and $(1,m,1)$, 
taken together, imply the unique alignment $(1,3,1)$ for $\phi_{\rm sol}$
in Eq.\ref{Phias2}.

To understand the origin of the phase $\eta=2\pi/3$ we shall start by imposing exact CP invariance on the
high energy theory, in Eqs.\ref{LS3} and \ref{l},
then spontaneously break CP in a very particular way, governed by the $Z_3$ symmetry,
so that $\eta$ is restricted to be a cube root of unity.
The Majoron flavon VEVs are driven by the superpotential,
\be
W_{S_4}^{\mathrm{flav,maj}} = P\left(\frac{\xi_{\rm atm}^3}{\Lambda}  -M^2\right) + 
P' \left(\frac{\xi_{\rm sol}^3}{\Lambda'}  - M'^2\right) ,
\label{Rflavon}
\ee
where $P,P'$ are two copies of ``driving'' superfields with $R=2$ but transforming as singlets under all other symmetries,
and 
$M$ is real due to CP conservation. Due to F-flatness,
\begin{equation}
 \left| \frac{\langle \xi_{\rm atm} \rangle^3}{\Lambda} - M^2\right|^2 
 = \left| \frac{\langle \xi_{\rm sol} \rangle^3}{\Lambda'} - M'^2\right|^2 = 0 .
\end{equation} 
These are satisfied by $\langle \xi_{\rm atm} \rangle = |(\Lambda M^2)^{1/3}|$ and 
$\langle \xi_{\rm sol} \rangle =  |(\Lambda'M'^2)^{1/3}|e^{-2i\pi/3}$
where we arbitrarily select the phases to be 
zero and $-2\pi /3$ from amongst a discrete set of possible choices
in each case. More generally we require a phase difference of $2\pi /3$ since the overall phase is not physically
relevant, which would happen one in three times by chance.
In the basis where the right-handed neutrino masses are real and positive
this is equivalent to $\eta  = 2\pi /3$ in Eq.\ref{eq:mnu2}, as in the benchmark
model in Eq.\ref{eq:mnu2p}, due to the see-saw mechanism.


\section{Conclusion}
\label{conclusion}
The seesaw mechanism provides an elegant explanation of the smallness of neutrino masses.
However in general it is difficult to test the mechanism experimentally, since the right-handed Majorana masses
may have very large masses out of reach of high energy colliders. The heavy Majorana sector also introduces 
a new flavour sector, with yet more parameters, beyond those describing low energy neutrino physics.
This is of serious concern, since the seesaw mechanism may be our best bet for extending the Standard Model
to include neutrino masses.

Given that the seesaw mechanism is an 
elegant but practically untestable mechanism with a large number of parameters,
in this paper we have relied on theoretical desiderata such as naturalness, minimality and predictability to guide us towards what we call the ``Littlest Seesaw'' model which is essentially the two right-handed neutrino model
bundled together 
with further assumptions about the structure of the Yukawa couplings that we call CSD($n$).
Understandably one should be wary of such assumptions,
indeed such principles of naturalness and minimality without experimental guidance could well prove to be unreliable. 
However we are encouraged by the fact that such principles in the guise of sequential dominance with 
a single texture zero, led to the bound $\theta_{13} \lesssim m_2/m_3$, suggesting a large reactor angle
a decade before it was measured. The additional CSD($n$) assumptions discussed here are
simply designed to explain
why this bound is saturated.

It is worth recapping the basic idea of sequential dominance that 
one of the right-handed neutrinos is dominantly responsible
for the atmospheric neutrino mass, while a subdominant 
right-handed neutrino accounts for the solar neutrino mass,
with possibly a third right-handed neutrino being approximately 
decoupled, leading to an effective two right-handed neutrino model. 
This simple idea leads to equally simple predictions which makes the scheme falsifiable.
Indeed, the litmus test of such sequential dominance
is Majorana neutrinos with a normal neutrino mass hiearchy and a very light (or massless)
neutrino. These predictions will be tested soon.

In order to understand why the reactor angle bound is approximately saturated,
we need to make additional assumptions, as mentioned. Ironically, the starting point is 
the original idea of constrained sequential dominance (CSD) which proved to be a good explanation
of the tri-bimaximal solar and atmospheric angles but 
predicted a zero reactor angle. However, this idea can be generalised to
the ``Littlest Seesaw'' comprising
a two right-handed neutrino model with constrained Yukawa
couplings of a particular CSD($n$) structure, where here $n>1$ is taken to be a real parameter.
We have shown that the reactor angle is given by 
$ \theta_{13} \sim (n-1) \frac{\sqrt{2}}{3}  \frac{m_2}{m_3}$
so that $n=1$ coresponds to original CSD with $ \theta_{13}=0$,
while $n=3$ corresponds to CSD($3$) with $ \theta_{13} \sim 2\frac{\sqrt{2}}{3}  \frac{m_2}{m_3}$,
corresponding to $\theta_{13} \sim m_2/m_3$, which provides an explanation for why the SD bound
is saturated as observed for this case, with both the approximation and SD breaking down for large $n$.

In general, the Littlest Seesaw is able to give a successful
desciption of neutrino mass and the PMNS matrix in terms of four input parameters appearing
in Eq.\ref{eq:mnu2} where the reactor angle requires $n\approx 3$. 
It predicts a normally ordered and very hierarchical neutrino
mass spectrum with the lightest neutrino mass being zero.
It also predicts TM1 mixing with the atmospheric sum rules providing further tests of the scheme.
Interestingly the single input phase $\eta$ 
must be responsible for \CP violation in 
both neutrino oscillations and leptogenesis, providing the most direct link
possible between these two phenomena. Indeed $\eta$ is identified as the leptogenesis phase.
Another input parameter is $m_b$ which is identied
with the neutrinoless double beta decay observable $m_{ee}$, although this is practically impossible to
measure for $m_1=0$.

The main {\em conceptual} achievement in this paper is to realise that making $n$ continuous greatly
simplifies the task of motivating the CSD($n$) pattern of couplings, which emerge almost as simply
as the TB couplings, as explained in Fig.\ref{perp}. 
The main {\em technical} achievement of the paper
is to provide exact analytic formulae for the lepton mixing angles, neutrino masses and CP phases
in terms of the four input parameters of CSD($n$) for any real $n>1$. The exact analytic results should facilitate phenomenological
studies of the LS model. We have checked our analytic results against the numerical
bechmark and validated them within the numerical precision. We also provided new simple analytic
approximations such as: $ \theta_{13} \sim (n-1) \frac{\sqrt{2}}{3}  \frac{m_2}{m_3}$ 
where $m_3\approx 2m_a$ and $m_2\approx 3m_b$.
The main {\em model building} achievement is to realise that the successful benchmark LS model 
based on CSD($3$) is quite well motivated by a discrete $S_4$ symmetry, 
since the neutrino vacuum alignment directions are enforced by residual symmetries that are contained in
$S_4$, but not $A_4$, which has hitherto been widely used in CSD($n$) models. 
This is illustrated by the starfish diagram in Fig,\ref{starfish}. In order to also fix the input leptogenesis phase to
its benchmark value $\eta=2\pi/3$, we proposed a benchmark model, including supersymmetric vacuum
alignment, based on $S_4\times Z_3\times Z'_3$,
which represents the simplest predictive seesaw 
model in the literature. The resulting benchmark predictions are:
solar angle $\theta_{12}=34^\circ$,  reactor angle $\theta_{13}=8.7^\circ$,  
atmospheric angle $\theta_{23}=46^\circ$, and 
Dirac phase $\delta_{CP}=-87^{\circ}$. These predictions are all within the scope of 
future neutrino facilities, and may provide a useful target for them to aim at.

\vspace{0.1in}
SFK acknowledges partial support from the STFC Consolidated ST/J000396/1 grant and 
the European Union FP7 ITN-INVISIBLES (Marie Curie Actions, PITN-
GA-2011-289442). 
\vspace{0.1in}

\appendix

\section{Lepton Mixing Conventions}
\label{conventions}
In the convention where the effective Lagrangian is given by
\footnote{Note that this convention for the 
light effective Majorana neutrino mass matrix $m^{\nu}$
differs by an overall complex conjugation compared to some other 
conventions in the literature.}
\begin{eqnarray}
{\cal L}=-  \overline{E_L} m^{E} E_R 
- \frac{1}{2}\overline{\nu_L} m^{\nu} \nu_{L}^c 
+ {H.c.}\; .
\label{Leff}
\end{eqnarray}
Performing the transformation from the flavour basis to the real positive mass basis by,
 \begin{eqnarray}\label{eq:DiagMe}
V_{E_L} \, m^E \,
V^\dagger_{E_R} = m^E_{\rm diag}=
\mbox{diag}(m_e,m_\mu,m_\tau)
 , \quad~
V_{\nu_L} \,m^{\nu}\,V^T_{\nu_L} = m^{\nu}_{\rm diag}=
\mbox{diag}(m_1,m_2,m_3),
\label{mLLnu}
\end{eqnarray}
the PMNS matrix is given by
\begin{eqnarray}\label{Eq:PMNS_Definition}
U
= V_{E_{L}} V^\dagger_{\nu_{L}} \,.
\end{eqnarray}
Since we are in the basis where the charged lepton mass matrix
$m^E$ is already diagonal, then in general $V_{E_{L}}$ can only be a diagonal matrix,
\beq
V_{E_{L}}=P_{E}=\left( \begin{array}{ccc}
e^{i\phi_{e}} & 0  & 0   \\
0 & e^{i\phi_{\mu}}  & 0 \\
0 & 0 & e^{i\phi_{\tau}}
\end{array}
\right),
\label{PE}
\eeq
consisting of arbitrary phases, where
an identical phase rotation on the right-handed charged leptons 
$V_{E_{R}}=P_{E}$ leaves the diagonal charged lepton masses in $m^E$ unchanged.
In practice the phases in $P_{E}$ are chosen to absorb three phases 
from the unitary matrix 
$V^\dagger_{\nu_{L}}$ and to put $U$ in a standard convention \cite{pdg},
\beq
U =V P
\label{DM}
\eeq
where, analogous to the CKM matrix,
\bea
 \label{eq:matrix}
V = 
\left(\begin{array}{ccc}
    c_{12} c_{13}
    & s_{12} c_{13}
    & s_{13} e^{-i\delta}
    \\
    - s_{12} c_{23} - c_{12} s_{13} s_{23} e^{i\delta}
    & \hphantom{+} c_{12} c_{23} - s_{12} s_{13} s_{23}
    e^{i\delta}
    & c_{13} s_{23} \hspace*{5.5mm}
    \\
    \hphantom{+} s_{12} s_{23} - c_{12} s_{13} c_{23} e^{i\delta}
    & - c_{12} s_{23} - s_{12} s_{13} c_{23} e^{i\delta}
    & c_{13} c_{23} \hspace*{5.5mm}
    \end{array}\right),
    \label{V}
\eea
and the Majorana phase matrix factor is, 
\beq
P  =
\label{Maj}
\left(\begin{array}{ccc}
e^{i\frac{\beta_1}{2}} & 0 & 0 \\
0 & e^{i\frac{\beta_2}{2}} & 0\\
0 & 0 & 1 \\
\end{array}\right).
\eeq
From Eqs.\ref{mLLnu},\ref{Eq:PMNS_Definition},\ref{PE}, we find,
\beq
U^{\dagger}P_E m^{\nu} P_E U^{*}=\mbox{diag}(m_1,m_2,m_3).
\eeq

\section{$S_4$}
\label{S4}
The irreducible representations of $S_4$ are two singlets $\bf{1}$ and $\bf{1'}$, one doublet $\bf{2}$
and two triplets $\bf{3}$ and $\bf{3'}$ \cite{Ishimori:2010au}.
The triplet $\bf{3}$ in the basis of \cite{Ishimori:2010au} corresponds to the following 24 matrices,
\begin{equation}
a_1=
\left( \begin{array}{ccc}
1 & 0 & 0\\
0 & 1 & 0 \\
0 & 0 & 1  
\end{array}
\right),\ \ 
a_2=
\left( \begin{array}{ccc}
1 & 0 & 0\\
0 & -1 & 0 \\
0 & 0 & -1  
\end{array}
\right),\ \ 
a_3=
\left( \begin{array}{ccc}
-1 & 0 & 0\\
0 & 1 & 0 \\
0 & 0 & -1  
\end{array}
\right),\ \ 
a_4=
\left( \begin{array}{ccc}
-1 & 0 & 0\\
0 & -1 & 0 \\
0 & 0 & 1  
\end{array}
\right)
\nonumber
\end{equation}
\begin{equation}
b_1=
\left( \begin{array}{ccc}
0 & 0 & 1\\
1 & 0 & 0 \\
0 & 1 & 0  
\end{array}
\right),\ \ 
b_2=
\left( \begin{array}{ccc}
0 & 0 & 1\\
-1 & 0 & 0 \\
0 & -1 & 0  
\end{array}
\right),\ \ 
b_3=
\left( \begin{array}{ccc}
0 & 0 & -1\\
1 & 0 & 0 \\
0 & -1 & 0   
\end{array}
\right),\ \ 
b_4=
\left( \begin{array}{ccc}
0 & 0 & -1\\
-1 & 0 & 0 \\
0 & 1 & 0  
\end{array}
\right)
\nonumber
\end{equation}
\begin{equation}
c_1=
\left( \begin{array}{ccc}
0 & 1 & 0\\
0 & 0 & 1 \\
1 & 0 & 0  
\end{array}
\right),\ \ 
c_2=
\left( \begin{array}{ccc}
0 & 1 & 0\\
0 & 0 & -1 \\
-1 & 0 & 0  
\end{array}
\right),\ \ 
c_3=
\left( \begin{array}{ccc}
0 & -1 & 0\\
0 & 0 & 1 \\
-1 & 0 & 0  
\end{array}
\right),\ \ 
c_4=
\left( \begin{array}{ccc}
0 & -1 & 0\\
0 & 0 & -1 \\
1 & 0 & 0  
\end{array}
\right)
\nonumber
\end{equation}
\begin{equation}
d_1=
\left( \begin{array}{ccc}
1 & 0 & 0\\
0 & 0 & 1 \\
0 & 1 & 0  
\end{array}
\right),\ \ 
d_2=
\left( \begin{array}{ccc}
1 & 0 & 0\\
0 & 0 & -1 \\
0 & -1 & 0  
\end{array}
\right),\ \ 
d_3=
\left( \begin{array}{ccc}
-1 & 0 & 0\\
0 & 0 & 1 \\
0 & -1 & 0  
\end{array}
\right),\ \ 
d_4=
\left( \begin{array}{ccc}
-1 & 0 & 0\\
0 & 0 & -1 \\
0 & 1 & 0  
\end{array}
\right)
\nonumber
\end{equation}
\begin{equation}
e_1=
\left( \begin{array}{ccc}
0 & 1 & 0\\
1 & 0 & 0 \\
0 & 0 & 1  
\end{array}
\right),\ \ 
e_2=
\left( \begin{array}{ccc}
0 & 1 & 0\\
-1 & 0 & 0 \\
0 & 0 & -1  
\end{array}
\right),\ \ 
e_3=
\left( \begin{array}{ccc}
0 & -1 & 0\\
1 & 0 & 0 \\
0 & 0 & -1  
\end{array}
\right),\ \ 
e_4=
\left( \begin{array}{ccc}
0 & -1 & 0\\
-1 & 0 & 0 \\
0 & 0 & 1  
\end{array}
\right)
\nonumber
\end{equation}
\begin{equation}
f_1=
\left( \begin{array}{ccc}
0 & 0 & 1\\
0 & 1 & 0 \\
1 & 0 & 0  
\end{array}
\right),\ \ 
f_2=
\left( \begin{array}{ccc}
0 & 0 & 1\\
0 & -1 & 0 \\
-1 & 0 & 0  
\end{array}
\right),\ \ 
f_3=
\left( \begin{array}{ccc}
0 & 0 & -1\\
0 & 1 & 0 \\
-1 & 0 & 0  
\end{array}
\right),\ \ 
f_4=
\left( \begin{array}{ccc}
0 & 0 & -1\\
0 & -1 & 0 \\
1 & 0 & 0  \end{array}
\right)
\label{3}
\end{equation}
where $a_i,b_i,c_i$ are the 12 matrices of the $A_4$ triplet representation,
while the remaining 12 matrices $d_i,e_i,f_i$ are the extra matrices in $S_4$.
The triplet $\bf{3'}$ in the basis of \cite{Ishimori:2010au} 
corresponds to matrices which are are simply related to those above,
\bea
a'_1=a_1, \ \ a'_2=a_2, \ \ a'_3=a_3, \ \ a'_4=a_4, \nonumber \\
b'_1=b_1, \ \ b'_2=b_2, \ \ b'_3=b_3, \ \ b'_4=b_4, \nonumber \\
c'_1=c_1, \ \ c'_2=c_2, \ \ c'_3=c_3, \ \ c'_4=c_4, \nonumber \\
d'_1=-d_1, \ \ d'_2=-d_2, \ \ d'_3=-d_3, \ \ d'_4=-d_4, \nonumber \\
e'_1=-e_1, \ \ e'_2=-e_2, \ \ e'_3=-e_3, \ \ e'_4=-e_4, \nonumber \\
f'_1=-f_1, \ \ f'_2=-f_2, \ \ f'_3=-f_3, \ \ f'_4=-f_4.
\label{3p}
\eea
In other words, for the $\bf{3'}$, the 12 matrices $a'_i,b'_i,c'_i$ associated with those of $A_4$ 
do not change sign,
while the remaining 12 matrices $d'_i,e'_i,f'_i$ involve a change of sign relative to the .

The Kronecker products of $S_4$ are: 
$\bf{1}\times \bf{1}=\bf{1}$, \ \ $\bf{1'}\times \bf{1'}=\bf{1}$, \ \ $\bf{1'}\times \bf{1}=\bf{1'}$,\ \ 
$\bf{3}\times \bf{3}=\bf{1}+\bf{2}+\bf{3}+\bf{3'}$, \ \ 
$\bf{3'}\times \bf{3'}=\bf{1}+\bf{2}+\bf{3}+\bf{3'}$, \ \ 
$\bf{3}\times \bf{3'}=\bf{1'}+\bf{2}+\bf{3}+\bf{3'}$, \ \ 
$\bf{2}\times \bf{2}=\bf{1}+\bf{1'}+\bf{2}$, \ \ 
$\bf{2}\times \bf{3}=\bf{3}+\bf{3'}$, \ \ 
$\bf{2}\times \bf{3'}=\bf{3}+\bf{3'}$.
The Clebsch relations in this basis are given in  \cite{Ishimori:2010au}.

\end{document}